\pdfoutput=1
\documentclass[iop]{emulateapj}

\slugcomment{}

\shorttitle{Physical environment of star-forming region W42}
\shortauthors{L.~K. Dewangan et al.}

\begin{document}

\title{Physical environment of massive star-forming region W42}
%{W42 complex: magnetic field, filaments, and star formation activity}
%\title[Physical environment of star-forming region W42]
\author{L.~K. Dewangan\altaffilmark{1}, A. Luna\altaffilmark{1}, D.~K. Ojha\altaffilmark{2}, B.~G. Anandarao\altaffilmark{3}, K.~K. Mallick\altaffilmark{2},\\ 
and Y.~D. Mayya\altaffilmark{1}}

\email{lokeshd@inaoep.mx}

\altaffiltext{1}{Instituto Nacional de Astrofísica, \'{O}ptica y Electr\'{o}nica, Luis Enrique Erro \# 1, Tonantzintla, Puebla, M\'{e}xico C.P. 72840}
\altaffiltext{2}{Department of Astronomy and Astrophysics, Tata Institute of Fundamental Research, Homi Bhabha Road, Mumbai 400 005, India}
\altaffiltext{3}{Physical Research Laboratory, Navrangpura, Ahmedabad - 380 009, India}

\begin{abstract}
We present an analysis of multi-wavelength observations from various datasets and Galactic plane surveys 
to study the star formation process in the W42 complex. 
A bipolar appearance of W42 complex is evident due to the ionizing feedback from the  O5-O6 type 
star in a medium that is highly inhomogeneous. 
The VLT/NACO adaptive-optics K and L$^{\prime}$ images (resolutions~$\sim$0\farcs2--0\farcs1) 
resolved this ionizing source into multiple point-like sources below $\sim$5000~AU scale.
The position angle $\sim$15$\degr$ of W42 molecular cloud is consistent with the H-band starlight mean polarization angle 
which in turn is close to the Galactic magnetic field, suggesting the influence of Galactic field on the evolution of the W42 molecular cloud.
{\it Herschel} sub-millimeter data analysis reveals three clumps located along the waist axis of the bipolar nebula, 
with the peak column densities of $\sim$3--5~$\times$~10$^{22}$ cm$^{-2}$ corresponding to visual extinctions of A$_{V}$ $\sim$32--53.5 mag.
The {\it Herschel} temperature map traces a temperature gradient in W42, revealing regions of 20~K, 25~K, and 30--36~K. 
{\it Herschel} maps reveal embedded filaments (length $\sim$1--3 pc) which appear to be radially pointed to the denser clump associated with the O5-O6 star, 
forming a hub-filament system. 
512 candidate young stellar objects (YSOs) are identified in the complex, $\sim$40\% of which are present in clusters distributed 
mainly within the molecular cloud including the {\it Herschel} filaments. 
Our datasets suggest that the YSO clusters including the massive stars are located at the junction of the filaments, similar to those seen in Rosette Molecular Cloud. 
\end{abstract}
\section{Introduction}
\label{sec:intro}
Active star-forming regions in molecular clouds often contain young star clusters, filaments, bubble(s), and massive star(s). 
Such regions are very promising sites for investigation to understand the formation and evolution of stellar clusters 
and the interaction between parent molecular cloud and embedded massive stars.

W42 is known as an obscured Galactic giant H\,{\sc ii} region (G25.38$-$0.18) towards the inner Galaxy \citep[e.g.][]{woodward85} and 
contains the IRAS 18355$-$0650 source. 
\citet{blum00} reported a foreground extinction of A$_{V}$ $\sim$10 mag in the direction of W42.  
\citet{woodward85} found a bipolar H\,{\sc ii} region with an angular size of $\sim$$12\farcs8$ ($\sim$0.24 pc at a distance of 3.8 kpc) and classified its ionizing source as an O7 star. 
The G25.38$-$0.18 H\,{\sc ii} region was further characterized as a core-halo structure with an angular size of $\sim$3$\arcmin$ 
($\sim$3.3 pc) \citep{garay93}. 
Recently, {\it Spitzer} images (3.6--24 $\mu$m) revealed that W42 has a bipolar appearance on a much 
larger scale \citep[see Figure~10 given in][]{deharveng10}, extending to $\sim3\farcm87$ ($\sim$4.3 pc), 
and is also known N39 \citep[e.g.,][]{beaumont10,churchwell06,deharveng10}. 
The entire complex, which includes an extended bipolar nebula associated with G25.38$-$0.18 H\,{\sc ii} region, 
is referred to as W42 complex in this work. 
\citet{blum00} investigated a small embedded stellar cluster at the heart of W42 using 
high spatial resolution near-infrared (NIR) images (hereafter, NIR cluster). They studied the NIR K-band spectra of three of the 
brightest stars in the cluster and one of the sources was identified as an O5-O6 star. It is thought that the W42 complex is most likely ionized by this star alone. 
However, the inner circumstellar environment of this source has not been studied.  
The $^{13}$CO profile along the line of sight to W42 shows two well-separated velocity components; one at 58--69 km s$^{-1}$ that is 
physically associated with W42, and the other at 88-109 km s$^{-1}$, that is associated with G25.4NW \citep[e.g.][]{ai13} 
located at the northwest edge of the extended bipolar nebula \citep[e.g.,][]{deharveng10}. 
Different velocity values suggest that these regions are not physically associated and are not part of the 
same star-forming complex \citep[e.g.,][]{ai13}. 
Using the C\,{\sc ii} and $^{3}$He radio recombination lines, \citet{quireza06} measured the velocity of the ionized gas to be 
$\sim$59.6 km s$^{-1}$ in W42. 
In the W42 complex, the velocities of $^{13}$CO gas and the ionized gas are consistent with a velocity of $\sim$61 km s$^{-1}$ obtained by 
ammonia gas (NH$_{3}$) \citep{wienen12}. 
\citet{blum00} reported a distance of $\sim$2.2 kpc to W42, assuming that the O-star is on the zero-age main sequence (ZAMS). 
\citet{anderson09} studied the properties of molecular cloud associated with the Galactic H\,{\sc ii} regions including W42 
(referred as U25.38$-$0.18 in their catalog). They assigned the $^{13}$CO a line center velocity of $\sim$64.61 km s$^{-1}$ 
that corresponds to a distance of $\sim$3.8 kpc. 
The near kinematic distance reported in the literature  \citep[e.g.][]{lester85} agrees well with this value. 
In this work, we adopt a distance of 3.8 kpc to the entire W42 complex. 
However, we show that our final conclusions are not affected by the distance discrepancy. 

Using 870 $\mu$m and 8 $\mu$m images of W42 complex, \citet{deharveng10} reported filaments or sheet-like features along the waist of the bipolar nebula.  Additionally, they found two dust condensations in the 870 $\mu$m image (in the northern and southern parts) along the waist of the bipolar nebula, 
and one of them (the southern part) was seen as an infrared dark cloud (IRDC) in the 8 $\mu$m image \citep{deharveng10}.  
Methanol maser was detected at 6.7-GHz in the northern condensation, with a velocity of $\sim$58.1 km s$^{-1}$ \citep{szymczak12}.  
Based on the expansion of H\,{\sc ii} region in a filament, \citet{deharveng10} highlighted physical scenarios to explain the existence of the bipolar 
morphology in N39. 

These previous studies on W42, in general, reveal the presence of ongoing star formation, the massive O5-O6 type star and of filament-like features.  
The study of filaments, their role in star formation process, and the interaction of a massive star with its natal molecular cloud
are yet to be explored observationally in the W42 complex. 
A knowledge of the physical environment of W42 at small and large scales is very essential to probe the ongoing physical mechanisms. 
To understand the physical conditions in the complex, 
we have utilized multi-wavelength data covering radio through NIR wavelengths from numerous surveys 
(e.g. the Multi-Array Galactic Plane Imaging Survey \citep[MAGPIS;][]{helfand06}, 
the Coordinated Radio and Infrared Survey for High-Mass Star Formation \citep[CORNISH;][]{hoare12}, 
the Galactic Ring Survey \citep[GRS;][]{jackson06}, the APEX Telescope Large Area Survey of the Galaxy \citep[ATLASGAL;][]{schuller09}, 
the {\it Herschel} Infrared Galactic Plane Survey \citep[Hi-GAL,][]{molinari10}, 
the MIPS Inner Galactic Plane Survey \citep[MIPSGAL;][]{carey05}, 
the Galactic Legacy Infrared Mid-Plane Survey Extraordinaire \citep[GLIMPSE;][]{benjamin03}, 
the UKIRT Wide-field Infrared Survey for H2 \citep[UWISH2;][]{froebrich11}, 
the UKIRT NIR Galactic Plane Survey \citep[GPS;][]{lawrence07}, 
the Galactic Plane Infrared Polarization Survey \citep[GPIPS;][]{clemens12}, the ESO Very Large Telescope (VLT) archive, 
and the Two Micron All Sky Survey data \citep[2MASS;][]{skrutskie06}). 
The ESO-VLT archival adaptive-optics NIR images are used to trace the small scale environment (inner 5000 AU) of the O5-O6 star. 
All these surveys are utilized to explore the different components associated with the complex, 
viz., molecular gas, ionized emission, dust (warm and cold) emission, shocked emission, distribution of dust temperature, column density, extinction, magnetic field, and embedded young stellar objects (YSOs).

This paper is organized as follows. In Section~\ref{sec:obser}, we describe in detail 
various datasets along with reduction procedures. Section~\ref{sec:data} focuses on the results 
related to the physical environment and point-like sources (classification and their analysis) 
of the region from various datasets. The possible star formation scenarios are discussed in 
Section~\ref{sec:data1}. Conclusions are presented in Section~\ref{sec:conc}.
\section{Data and analysis}
\label{sec:obser}
We have used multi-wavelength data to study the star formation and feedback of massive star on the surrounding 
interstellar medium (ISM) in the W42 complex. 
The size of the selected area is $\sim15\farcm4  \times 14\farcm4$, centered 
at $\alpha_{2000}$ = 18$^{h}$38$^{m}$14$^{s}$, $\delta_{2000}$ = $-$06$\degr$47$\arcmin$24$\arcsec$, 
which corresponds to a physical scale of $\sim$17 pc $\times$ 16 pc.
\subsection{NIR Data}
NIR photometric data were obtained from the UKIDSS 6$^{th}$ archival data release (UKIDSSDR6plus) of the GPS. 
UKIDSS observations were performed using the UKIRT Wide Field Camera
\citep[WFCAM;][]{casali07}. The final fluxes were calibrated using the 2MASS data. 
A full description of data reduction and calibration procedures are given in
\citet{dye06} and \citet{hodgkin09}, respectively. 
In order to select only reliable point sources from the GPS catalog, we utilized the criteria 
suggested by \citet{lucas08}. For the sources common in all the three NIR ({\it JHK}) bands and 
those detected only in the {\it H} and {\it K} 
bands, separate criteria\footnote[1]{In the appendix-\ref{app:cond}, we give the SQL script that is especially written to implement these criteria.} 
were adopted for selection. 
These criteria include the removal of saturated sources, non-stellar
sources, and unreliable sources near the sensitivity limits. 
Following the criteria, our selected GPS catalog contains sources fainter than J = 12.5, H = 11.6, and K = 10.2 mag to avoid saturation. 
The magnitudes of saturated bright sources were retrieved from the 2MASS catalog. 
We selected only those sources which have magnitude error of 0.1 
or less in each band, to obtain good photometric quality. 
Following the above criteria, we found 17745 sources common to all the three J, H, and K bands. 
In addition to these detections, 8754 sources having only the H and K bands detection were also obtained.

We retrieved archival adaptive-optics imaging data towards W42 region from the ESO-Science Archive 
Facility (ESO proposal ID: 089.C-0455(A); PI: Jo\~{a}o Alves). 
Observations were made with 8.2m VLT with 
NAOS-CONICA (NACO) adaptive-optics system \citep{lenzen03,rousset03} 
in K$_{s}$-band ($\lambda _{c}=2.18\, \mu \rm m, \Delta \lambda =0.35\, \mu \rm m$) and 
L$^{\prime}$-band ($\lambda _{c}=3.80\, \mu \rm m, \Delta \lambda =0.62\, \mu \rm m$). 
We obtained five K$_{s}$ frames and six L$^{\prime}$ frames of 24 and 21 seconds of exposures, respectively. 
The final processed NACO images were obtained through the standard analysis procedure such as sky subtraction, 
image registration, combining with median method, and astrometric calibration, using IRAF and STAR-LINK softwares. 
The astrometry calibration of NACO images was performed using the GPS K-band point sources.
NACO K$_{s}$ and L$^{\prime}$ images have plate scales of 0\farcs054/pixel and 0\farcs027/pixel, respectively, 
with resolutions varying from 0\farcs2 ($\sim$760 AU) -- 0\farcs1 ($\sim$380 AU). 
\subsection{H$_{2}$ Narrow-band Image}
Narrow-band H$_{2}$ (v = $1-0$ S(1)) image at 2.12 $\mu$m was retrieved from the UWISH2 database. 
To obtain a continuum-subtracted H$_{2}$ map, the point spread function of GPS K-band image was matched 
and scaled to the H$_{2}$ image.
\subsection{H-band Polarimetry}
Imaging polarimetry in NIR H-band (1.6 $\mu$m) was obtained from the GPIPS\footnote[2]{http://gpips0.bu.edu/}. 
Observations were taken with the {\it Mimir} instrument, on the 1.8 m Perkins telescope, 
in H-band linear imaging polarimetry mode \citep[see][for more details]{clemens12}.
The polarization data towards W42 complex were covered in two GPIPS fields i.e., 
GP0612 ($l$ = 25$\degr$.319, $b$ = $-$0$\degr$.240) and GP0626 ($l$ = 25$\degr$.447, $b$ = $-$0$\degr$.165). 
Only those sources with Usage Flag (UF) = 1 and $P/\sigma_p \ge$ 2 were selected for the study, to 
obtain good polarimetric quality, where $P$ is the polarization percentage and $\sigma_p$ is the polarimetric
uncertainty. These conditions provided a total of 234 stars from two GPIPS fields.  
\subsection{{\it Spitzer} Data}
{\it Spitzer} Infrared Array Camera \citep[IRAC;][]{fazio04} 3.6--8.0 $\mu$m
images were obtained from the GLIMPSE survey. We obtained the photometry of point sources from the GLIMPSE-I Spring '07 highly reliable catalog. 
The GLIMPSE-I catalog does not provide photometric magnitudes of some sources, which are well detected in the images. 
Aperture photometry was performed for such sources using the GLIMPSE images at a plate scale
of 0$\farcs$6/pixel. The photometry was extracted using a
2$\farcs$4 aperture radius and a sky annulus from 2$\farcs$4 to
7$\farcs$3 in IRAF\footnote[3]{IRAF is distributed by the National Optical Astronomy Observatory, USA}. 
Apparent magnitudes were calibrated using the IRAC zero-magnitudes 
(i.e., 18.59 (3.6 $\mu$m), 18.09 (4.5 $\mu$m), 17.49 (5.8 $\mu$m), and 16.70 (8.0 $\mu$m)) including aperture 
corrections \citep[see IRAC Instrument Handbook (Version 1.0, 2010 February) and also][]{dewangan12}. 
We also utilized MIPSGAL 24 $\mu$m photometry in this work. 
We performed aperture photometry on the MIPSGAL 24 $\mu$m image to extract point sources. 
The photometry was obtained using a $7\arcsec$ aperture radius and a sky annulus from $7\arcsec$ to $13\arcsec$ in IRAF \citep[e.g.][]{dewangan15}. 
MIPS zero-magnitude flux density, including aperture correction was used for the photometric calibration, as reported in the MIPS 
Instrument Handbook-Ver-3\footnote[4]{http://irsa.ipac.caltech.edu/data/SPITZER/docs/mips/mipsinstrumenthandbook/}. 

\subsection{{\it Herschel} and ATLASGAL Data}
{\it Herschel} Hi-GAL continuum maps 
were retrieved for 70 $\mu$m, 160 $\mu$m, 250 $\mu$m, 350 $\mu$m, and 500 $\mu$m. 
The beam sizes of these bands are 5$\farcs$8, 12$\arcsec$, 18$\arcsec$, 25$\arcsec$, and 37$\arcsec$ \citep{poglitsch10,griffin10}, respectively. 
We obtained the processed level2$_{-}$5 products, using the {\it Herschel} Interactive Processing 
Environment \citep[HIPE,][]{ott10}. The images at 70--160 $\mu$m were in the units of Jy pixel$^{-1}$, 
while the images at 250--500 $\mu$m were calibrated in the surface brightness unit of MJy sr$^{-1}$. 
The plate scales of the 70, 160, 250, 350, and 500 $\mu$m images are 3.2, 6.4, 6, 10, and 14 arcsec/pixel, respectively.

The sub-millimeter (mm) continuum map at 870 $\mu$m (beam size $\sim$19$\farcs$2) was obtained from the APEX ATLASGAL 
archival survey. 
\subsection{$^{13}$CO (J=1$-$0) Line Data}
The $^{13}$CO (J=1$-$0) line data were downloaded from the GRS\footnote[5]{http://www.bu.edu/galacticring/}. 
The survey data have a velocity resolution of 0.21~km\,s$^{-1}$, an angular resolution 
of 45$\arcsec$ with 22$\arcsec$ sampling, a typical rms sensitivity (1$\sigma$)
of $\approx0.13$~K, a velocity coverage of $-$5 to 135~km~s$^{-1}$, and a main 
beam efficiency ($\eta_{\rm mb}$) of $\sim$0.48 \citep{jackson06}.  
\subsection{Radio Centimeter Continuum Map}
Radio continuum map at 20 cm (beam size $\sim$6$\arcsec$) was obtained from the VLA MAGPIS\footnote[6]{http://third.ucllnl.org/gps/index.html}. 
CORNISH\footnote[7]{http://cornish.leeds.ac.uk/public/index.php} 5~GHz (6 cm) high-resolution radio continuum 
data (beam size $\sim$1\farcs5) are also utilized in this work. CORNISH 5 GHz compact source catalog 
was also retrieved from \citet{purcell13}. 
\subsection{Other Data}
We used the observed positions of the 6.7-GHz methanol maser \citep{szymczak12} 
and the O5-O6 star \citep{blum00}. 
The properties of the molecular cloud U25.38$-$0.18 \citep{anderson09} are also utilized in this work.
\section{Results}
\label{sec:data}
\subsection{Multi-phase environment of W42 complex}
\subsubsection{The bipolar nebula and its heart}
\label{subsec:data1}
\begin{figure*}
\epsscale{1.0}
\plotone{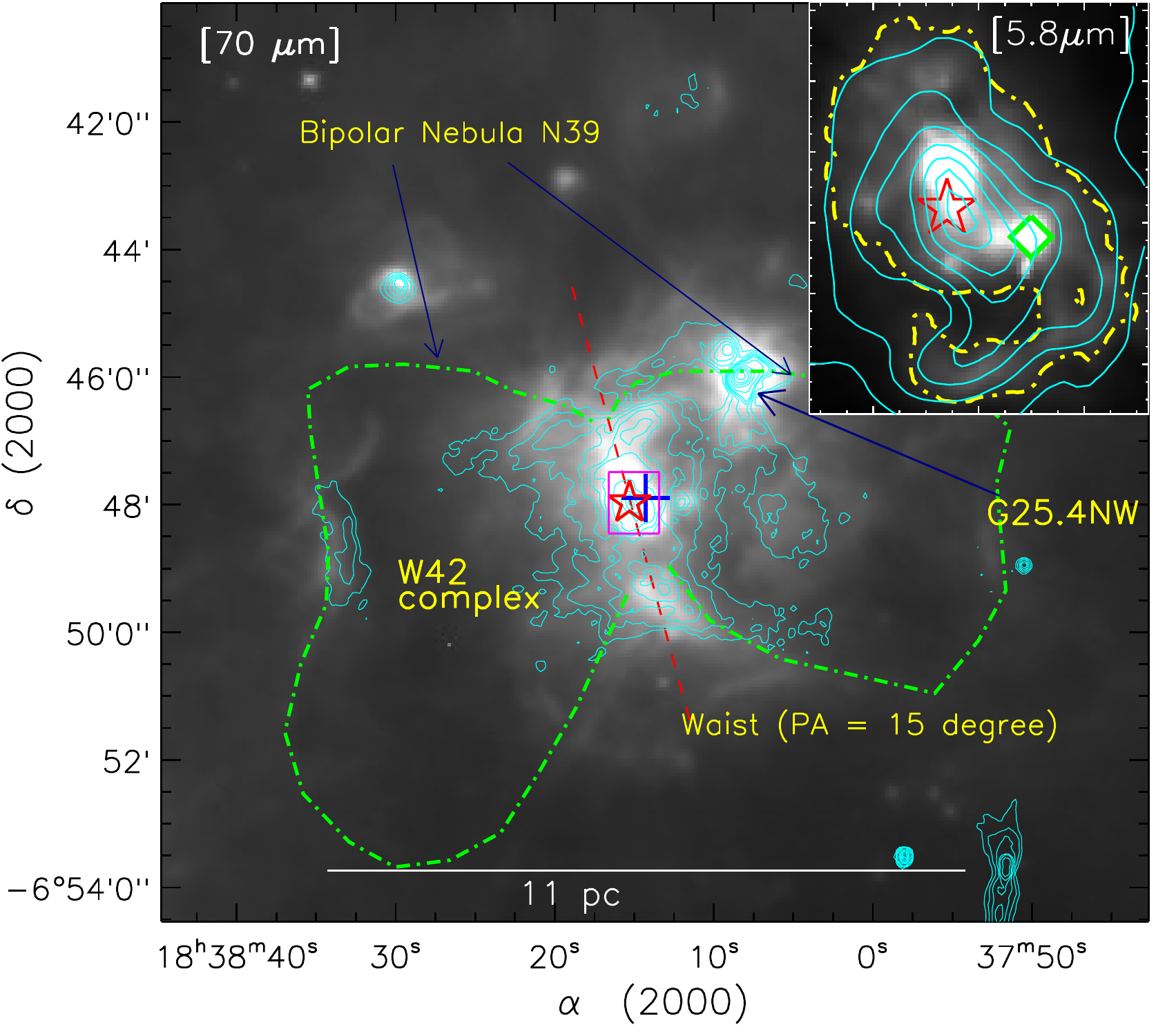}
\caption{\scriptsize {\it Herschel} 70 $\mu$m emission view of W42 complex (size of the selected field $\sim 15\farcm4  \times 14\farcm4$; central
coordinates: $\alpha_{2000}$ = 18$^{h}$38$^{m}$13$^{s}$.9, $\delta_{2000}$ = $-$06$\degr$47$\arcmin$24$\arcsec$.4). 
Contours of MAGPIS 20 cm radio emission in cyan color are superimposed with levels of 
0.003, 0.006, 0.011, 0.017, 0.024, 0.047, 0.094, 0.141, 0.189, 0.260, 0.331, 0.402, and 0.449 Jy/beam. 
The positions of IRAS 18355$-$0650 (+) and 
an O5-O6 star (star symbol) are marked in the figure. 
The bipolar nebula is depicted from the {\it Spitzer} 8 $\mu$m emission (dot-dashed green contour) \citep[see][]{deharveng10}. 
The scale bar on the bottom shows a size of 11 pc at a distance of 3.8 kpc. The magenta box is shown as a zoomed-in view in Figure~\ref{fig2}a.
The inset on the top right shows the central region in zoomed-in view, using {\it Spitzer}-IRAC 5.8 $\mu$m image in linear gray-scale 
(see the magenta box in the main figure). The 5.8 $\mu$m contour emission (dot-dashed yellow color) is shown with a level of 555 MJy/Sr. 
MAGPIS 20 cm emissions are also overlaid by cyan contours on 
the inset image with the same levels as shown in the main figure. 
The positions of a 6.7-GHz methanol maser ($\Diamond$) and an O5-O6 star (star symbol) are marked in the inset figure.}
\label{fig1}
\end{figure*}

The spatial distribution of warm dust emission traced in {\it Herschel} 70 $\mu$m image towards 
our selected region (size~$\sim15\farcm4  \times 14\farcm4$) 
is shown in Figure~\ref{fig1}, which depicts the previously known bipolar nebular morphology. 
The bubble region appears to extend several parsecs ($\sim$11 pc $\times$ 7 pc). 
As traced in 8.0 $\mu$m emission, the waist and the edges of the nebula are highlighted in Figure~\ref{fig1} 
\citep[also see Figure~10 in][]{deharveng10}. The 24 $\mu$m image also traces the warm dust emission and 
is saturated near the location of the O5-O6 star \citep[see Figure~10 given in][]{deharveng10}. 
Furthermore, the 24 $\mu$m emission is enclosed by the 8.0 $\mu$m emission. 
The G25.4NW region lies in the direction of the W42 complex, but is not associated with it, 
as can be seen from its very different velocity \citep{ai13} which indicates a very different distance. 
Therefore, the results of G25.4NW region are not discussed in this work. 
The radio continuum emission at MAGPIS 20 cm (beam $\sim$6$\arcsec$), which traces the ionized emission, shows 
the central $\sim$4.28 pc part of the bipolar nebula. The peak of 20 cm emission is found to 
be approximately coincident with a well-classified O5-O6 star (see Figure~\ref{fig1}). 
This source could be the powering source of the entire extended emission. 
Using the MAGPIS 20 cm continuum data, \citet{beaumont10} estimated the Lyman continuum photon 
number (logN$_{uv}$) within the N39 nebula to be $\sim$49.24 for a distance of 3.7 kpc (or logN$_{uv}$ $\sim$49.26 for a distance of 3.8 kpc). 
This observed N$_{uv}$ value is in agreement with the 
theoretical N$_{uv}$ value for a spectral type of O5 star \citep[logN$_{uv}$ $\sim$49.26;][]{martins05}. 
In this case, we assume that no ionizing photons are absorbed by the dust in the ionized gas, which is probably not true because of the presence of 
warm dust emission inside the H\,{\sc ii} region. As mentioned before, \citet{blum00} have reported a distance of 2.2 kpc to W42. 
We find logN$_{uv}$ $\sim$48.79 for a distance of 2.2 kpc, which corresponds to a single ionizing star of spectral type O6.5V (see Table 1 in \citet{martins05} for theoretical values). The radio spectral type of the exciting star at 2.2 and 3.8 kpc is consistent with a MK type spectrum of O5--O6 \citep[e.g.][]{blum00}. 

\begin{figure*}
\epsscale{1.0}
\plotone{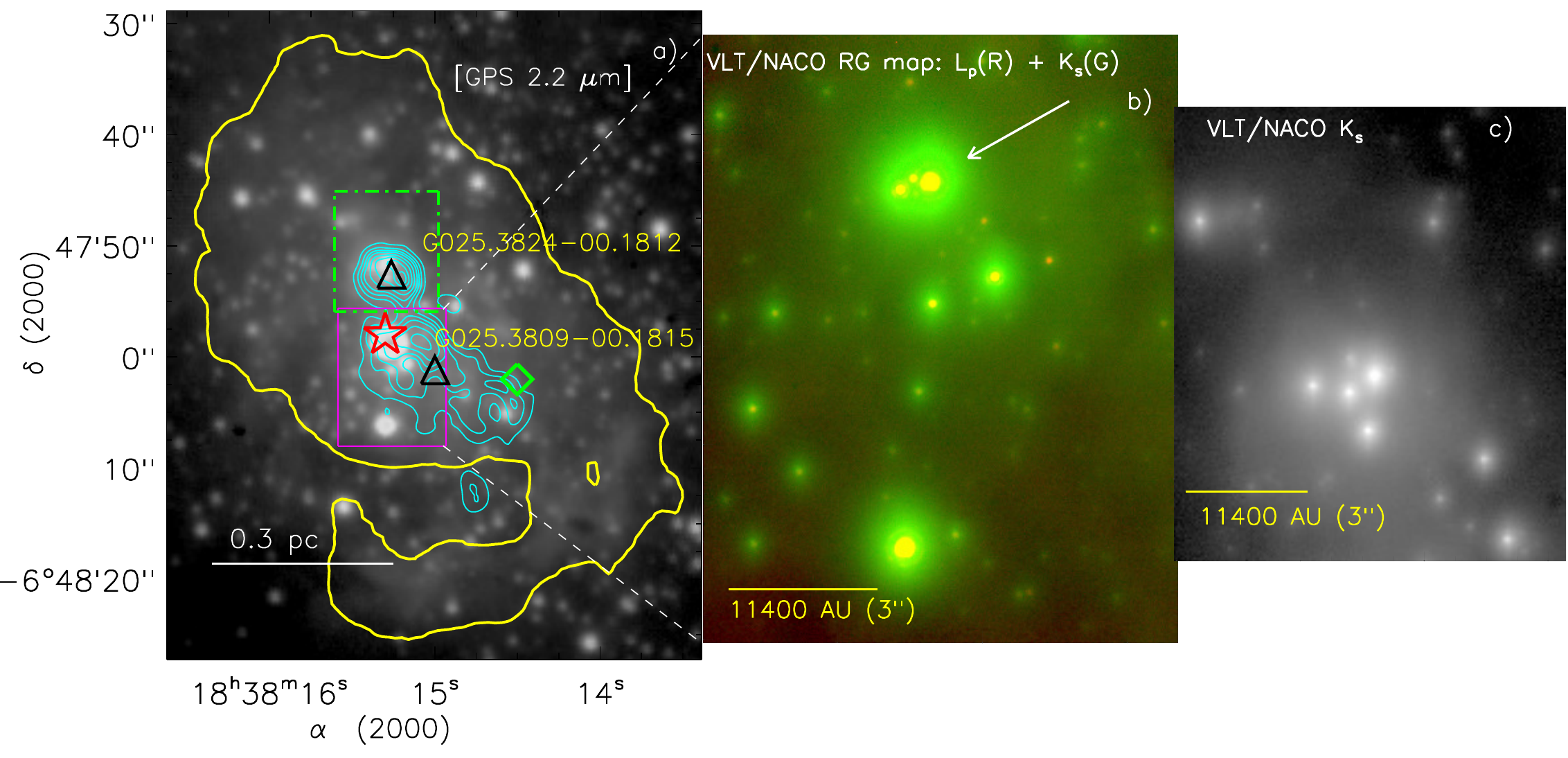}
\caption{\scriptsize a) Figure shows the central region in zoomed-in view as highlighted by a magenta box in Figure~\ref{fig1}, 
using GPS-K band image in logarithmic gray-scale. {\it Spitzer}-5.8 $\mu$m contour emission (in yellow color) is shown with a level of 555 MJy/Sr. 
CORNISH 5 GHz emissions (beam size $\sim$1\farcs5) are also overlaid by cyan contours on 
the image with the levels of 0.035 Jy/beam $\times$ (0.1, 0.2, 0.3, 0.4, 0.55, 0.7, 0.85, 0.95). 
The positions of a 6.7-GHz methanol maser emission ($\Diamond$) and an O5-O6 star ($\star$) are marked in the figure. 
The peak positions of two compact radio sources (i.e. G025.3809$-$00.1815 and G025.3824$-$00.1812) are shown by triangle symbols. 
A zoomed-in view of the solid magenta box is shown in Figure~\ref{fig2}b. 
A zoomed-in view of the dot-dashed green box is shown in Figure~\ref{fig2}c. 
b) VLT/NACO adaptive-optics K$_{s}$ and L$^{\prime}$ images (in logarithmic scale) towards G025.3809$-$00.1815. 
The position of an O5-O6 star is highlighted by a white arrow. 
c) VLT/NACO adaptive-optics K$_{s}$ image (in logarithmic gray scale) towards G025.3824$-$00.1812. 
Note that there is no VLT/NACO L$^{'}$ image available towards this radio source.} 
\label{fig2} 
\end{figure*}

The inset on the top right of Figure~\ref{fig1} shows the heart of the bipolar nebula, 
as seen in the {\it Spitzer}-IRAC 5.8 $\mu$m image overlaid with MAGPIS 20 cm and 
IRAC 5.8 $\mu$m contours. The 5.8 $\mu$m contour emission reveals a structure (a spatial extension of $\sim$1 pc scale), 
which was interpreted as an ionized cavity-like structure by \citet{dewangan15b}. 
This structure is well traced in H$_{2}$ 2.12 $\mu$m emission and in the continuum emission at 3.6--8.0 $\mu$m, and 20 cm \citep[see][]{dewangan15b}. 
CORNISH 5~GHz high-resolution radio continuum emission \citep[beam $\sim$1$\farcs$5;][]{purcell13} traces two 
compact radio sources (i.e. G025.3824$-$00.1812 (angular scale $\sim$3$\farcs$1) and G025.3809$-$00.1815 
(angular scale $\sim$8$\farcs$3)) located inside this cavity-like structure (see Figure~\ref{fig2}a). 
CORNISH 5~GHz radio contours are overlaid on the 
GPS K band image in Figure~\ref{fig2}a.
A comparison of radio emission between MAGPIS 20 cm (beam $\sim$6$\arcsec$) and CORNISH 5~GHz (beam $\sim$1\farcs5) 
can be seen in Figures~\ref{fig1} and~\ref{fig2}a. 
\citet{purcell13} also measured the integrated flux densities equal to 200.13 mJy and 460.83 mJy 
for G025.3824$-$00.1812 and G025.3809$-$00.1815 sources, respectively (see Table~\ref{tab1} for coordinates). 
The spectroscopically identified O5-O6 star is located within the extension of G025.3809$-$00.1815. 
The peak positions of these radio sources, G025.3824$-$00.1812 and G025.3809$-$00.1815, are $\sim$5$\farcs$4 
and $\sim$5$\farcs$5 away from the location of the O5-O6 star, respectively. 
In order to infer the spectral class of each of the compact radio sources, we have estimated the number of Lyman 
continuum photons (N$_{uv}$) using the integrated flux density following the equation of \citet{matsakis76} \citep[also see][for more details]{dewangan15}.
The calculations were carried out for a distance of 3.8 kpc and for the electron temperature of 10000~K.
We compute N$_{uv}$ (or logN$_{uv}$) to be $\sim$2.6 $\times$ 10$^{47}$ s$^{-1}$ (47.41) and $\sim$5.9 $\times$ 10$^{47}$ s$^{-1}$ (47.77) 
for G025.3824$-$00.1812 and G025.3809$-$00.1815, respectively. 
These values correspond to a single ionizing star of 
spectral type B0V (see Table 1 in \citet{smith02} for theoretical values) and O9.5V 
(see Table 1 in \citet{martins05} for theoretical values) for G025.3824$-$00.1812 and G025.3809$-$00.1815, respectively.  
These radio peaks are individual H\,{\sc ii} regions. Note that the extended radio emission at 20 cm does not appear to 
be produced by these B0V and O9.5V type stars together. There is no infrared counterpart to the peak of radio source 
G025.3809$-$00.1815 seen in the GPS NIR image, while a NIR counterpart of G025.3824$-$00.1812 is found in the GPS NIR image. 
Note that in the presence of the O5-O6 star, there will hardly be any noticeable additional effect of 
B0V spectral type star, because the radio flux of B0V type star ($\sim$0.2 Jy) is about 1/80$^{th}$ of the O5-O6 type star (i.e. $\sim$16 Jy). 
Therefore, the W42 complex appears more likely to be excited by the spectroscopically identified O5-O6 type star. 

In order to study the small scale environment of the O5-O6 type star, we utilized the ESO-VLT archival NIR images.
In Figure~\ref{fig2}b, we present the VLT/NACO adaptive-optics NIR images towards G025.3809$-$00.1815 (including the O5-O6 star) 
in K$_{s}$ and L$^{\prime}$ bands. 
Figure~\ref{fig2}c shows the VLT/NACO K$_{s}$ image towards G025.3824$-$00.1812 radio source. 
The VLT/NACO images provide the circumstellar view of the O5-O6 star in inner regions within 5000 AU. 
Within this scale, the O5-O6 star is resolved into at least two stellar sources in the K$_{s}$ image, 
while the L$^{\prime}$ image resolves the O5-O6 star into at least three stellar sources. 
We do not find any nebular feature in NACO images within 10000~AU of the O5-O6 star. 
In general, it is observationally known that massive stars are often found in binary and multiple systems \citep{duchene13}. 
In M8 massive star-forming region, \citet{goto06} resolved an ``O" star (designated as ``$Herschel-36$") into multiple sources using the VLT/NACO images. 
It appears that the O5-O6 star in W42 is associated with a very similar system of multiple stellar sources like $Herschel-36$ in M8 \citep[see Figure~1 in][]{goto06}.  
The VLT/NACO K$_{s}$ image shows at least 5 stellar sources towards the peak position of G025.3824$-$00.1812, within a scale of $\sim$10000 AU.
There are no nebular features seen towards these sources in NACO images. 
We do not have detections of these resolved sources in both NACO bands, 
hence we cannot provide any quantitative predictions (such as color and spectral type) in this work. 
Previously, \citet{blum00} detected only three point-like sources towards G025.3824$-$00.1812 and 
studied the K-band spectra of one of the three sources \citep[see source \#3 marked in Figure~1 in][]{blum00}. 
These authors did not detect any stellar absorption features, and also found NIR excess emission 
associated with source, which led them to suggest this source as a YSO candidate having extinction of about 10 mag. 
High resolution spectroscopic study will be helpful to identify main ionizing source of the complex. 
On the other hand, it is also probable that the W42 complex is being ionized by a small cluster of O and early B type stars 
rather than by a single star.
Additionally, \citet{dewangan15b} reported a parsec scale H$_{2}$ outflow that is driven by an infrared 
counterpart of the 6.7-GHz methanol maser emission, namely, W42-MME. 
W42-MME is observed at wavelengths longer than 2.2~$\mu$m and is classified as a deeply embedded massive YSO (stellar mass $\sim$19$\pm$4 M$_{\odot}$ and extinction $\sim$48$\pm$15 mag) (see Figure~\ref{fig2}a) \citep{dewangan15b}. 
They also investigated a jet-like feature in W42-MME using the VLT NIR adaptive-optics images.  
The O5-O6 star is located at a projected linear separation of about 0.22 pc from the W42-MME. 
These observational features suggest that massive star formation is currently taking place in the W42 complex (also see Section~\ref{subsec:sfm}). 

\subsubsection{Tracers: warm dust, cold dust, and molecular gas}
\label{subsec:data2}
\begin{figure*}
\epsscale{1.0}
\plotone{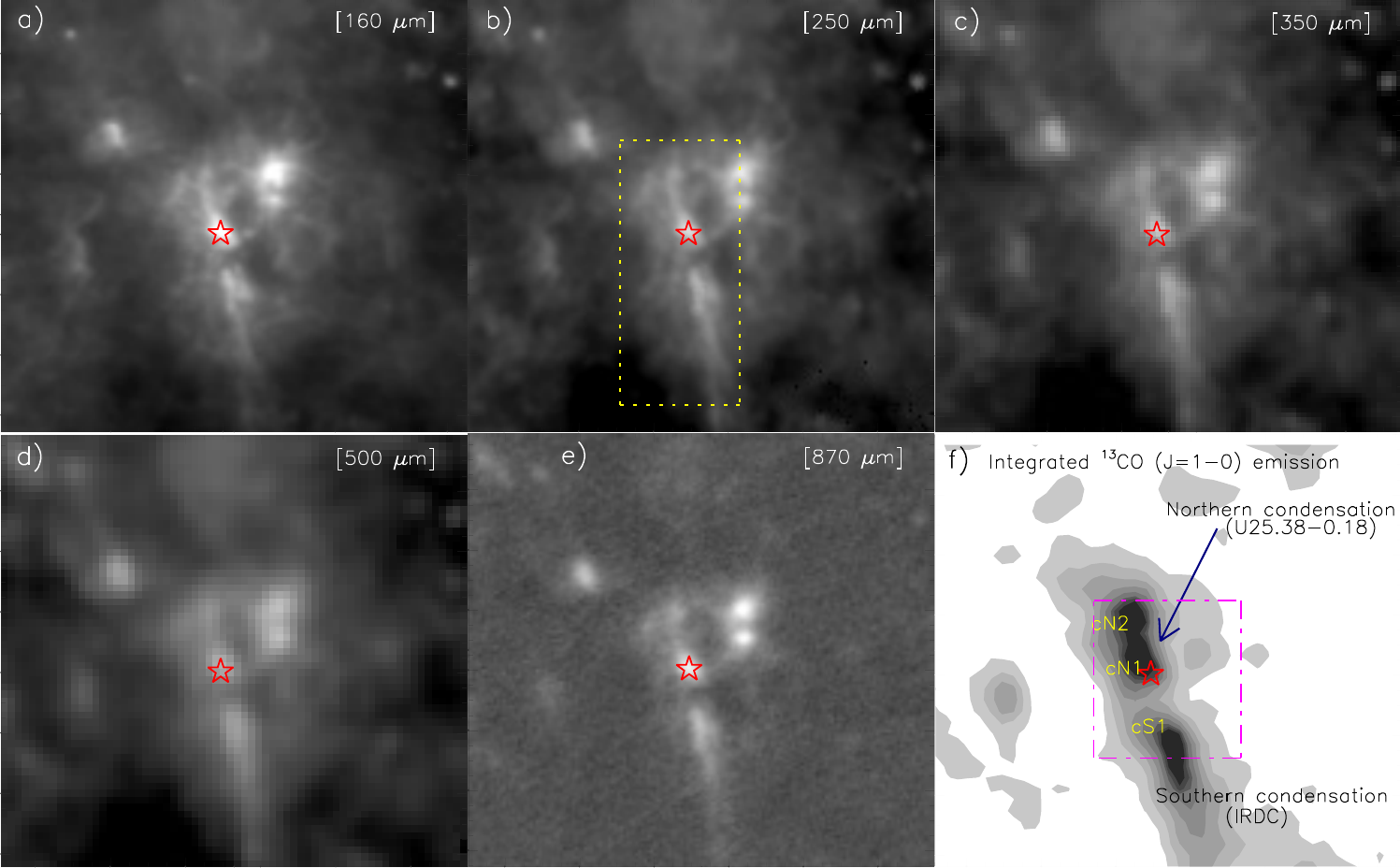}
\caption{\scriptsize {\it Herschel} dust and molecular emissions of W42 complex, using the Hi-GAL, ATLASGAL, and GRS surveys.
The panels show images at 160 $\mu$m, 250 $\mu$m, 350 $\mu$m, 500 $\mu$m, 
870 $\mu$m, and integrated $^{13}$CO contour map, from left to right in increasing order. 
In all the panels, the star symbol indicates the location of an O5-O6 star. 
A zoomed-in view of the dotted box in the panel ``b" is shown in Figure~\ref{fig4}a. 
The contour map of integrated $^{13}$CO emission is shown in the velocity range of 58--69 km s$^{-1}$. 
The CO contours are 2.37 K km s$^{-1}$ $\times$ (3, 5, 7, 9, 11, 13, 14, 15). 
The magenta box in the panel ``f" is shown as a zoomed-in view in Figure~\ref{fig4}b. 
Based on the visual inspection of the integrated CO map, two clumps in the northern 
condensation (i.e. cN1 and cN2) and one clump (cS1) in the southern condensation are marked in panel ``f". } 
\label{fig3} 
\end{figure*}

\begin{figure*}
\epsscale{1.0}
\plotone{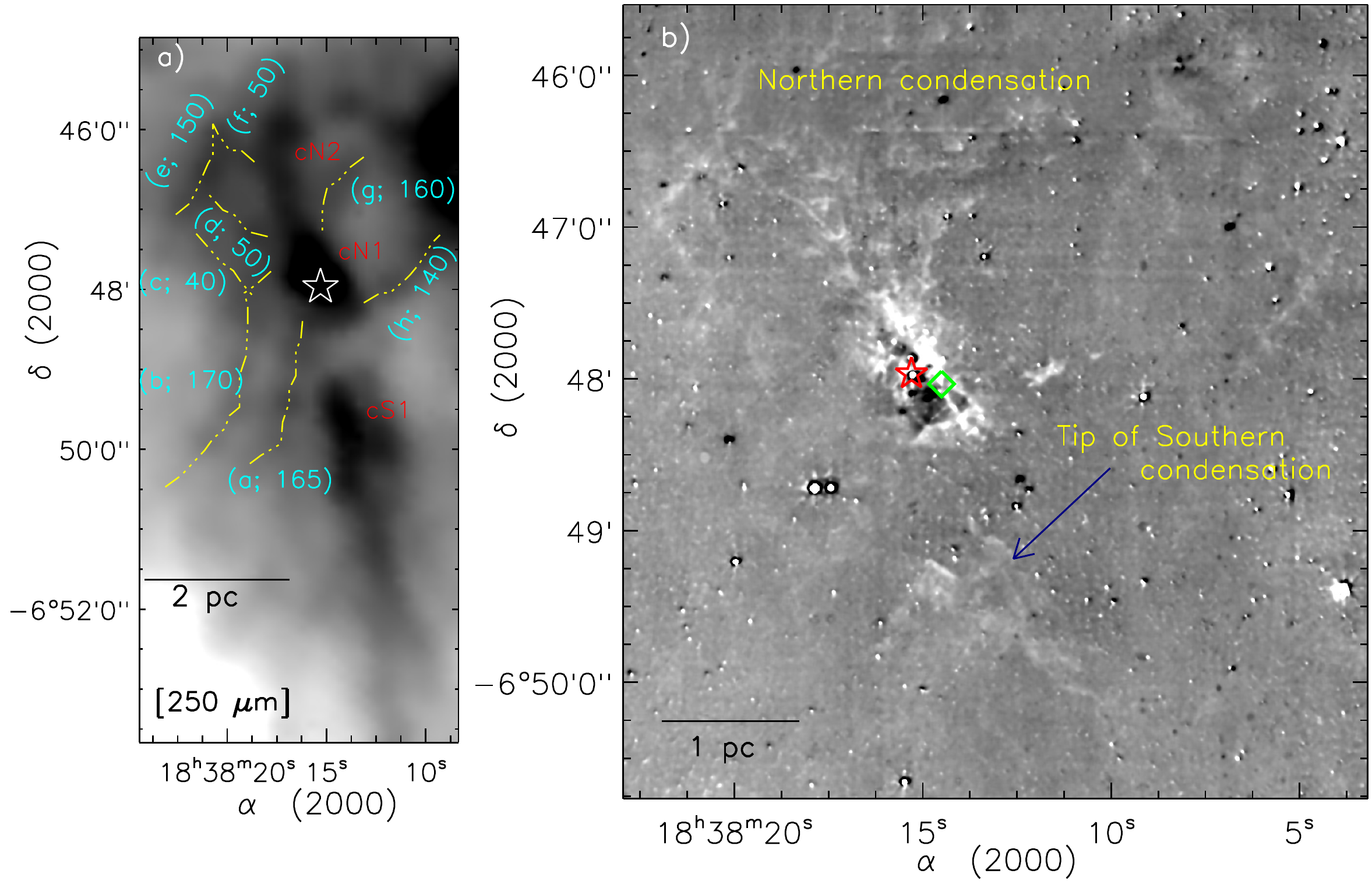}
\caption{\scriptsize a) Inverted gray scale map of the waist axis of the bipolar bubble using {\it Herschel} 250 $\mu$m image. 
The selected area of 250 $\mu$m image is shown by a box in Figure~\ref{fig3}b. 
Filaments are shown by yellow color curves along with their designations and position angles. 
b) Continuum-subtracted H$_{2}$ image (gray-scale) at 2.12 $\mu$m of the northern condensation (size of the region $\sim 4\farcm9 \times 5\farcm2$), as shown by a magenta box in Figure~\ref{fig3}f. The marked symbols are similar to those shown in Figure~\ref{fig2}a.}
\label{fig4}
\end{figure*}

In Figure~\ref{fig3}, we show a longer wavelength view of the W42 complex.
The images are shown at 160 $\mu$m, 250 $\mu$m, 350 $\mu$m, 500 $\mu$m, 870 $\mu$m, and integrated $^{13}$CO (J=1$-$0) intensity contour map. 
The CO map is integrated in the [58,69] km s$^{-1}$ velocity range. 
The 160--870 $\mu$m emission traces cold dust components (see Section~\ref{subsec:temp} for quantitative information). 
Most of the emission in all the maps is concentrated toward the waist of the bipolar nebula (i.e., two condensations, namely northern and southern), as previously reported by \citet{deharveng10}. Both of these condensations have a similar position angle of $\sim$15$\degr$. 
CO data are very useful to study the morphology of molecular cloud.  
In Figure~\ref{fig3}f, an elongated molecular cloud structure is evident along the waist of the nebula, where the northern and southern 
condensations are clearly visible as prominent bright features. 
The northern dust component appears to contain two CO clumps (cN1 and cN2) and is associated 
with previously known NIR cluster including the 6.7-GHz maser and the O5-O6 type star (see Table~\ref{tab1} for coordinates). 
The southern condensation, traced as a CO clump (cS1), is found to be linked with the IRDC seen at 8.0 $\mu$m.  
The IRDC appears as a bright filament (length $\sim$5 pc) in emission at wavelengths longer than 70 $\mu$m. 
In addition to the northern and southern condensations, the filamentary features (labels: a, b, c, d, e, f, and h) are also present in the Hi-GAL maps. 
In the proximity of the northern condensation, the filamentary features are highlighted based on visual inspection of the 
Hi-GAL 250 $\mu$m map (see curves in Figure~\ref{fig4}a; hereafter {\it Herschel} filaments). 
The position angles of these filaments ($\sim$40--170 degrees) are also shown in Figure~\ref{fig4}a. 
The lengths of these filaments vary between 1 and 3 pc scales. 
The orientations of these filaments are different from the position angles of the northern and southern condensations.  
Note that these {\it Herschel} filamentary features are not detected in the 870 $\mu$m map (see Figure~\ref{fig3}e).  

Taken together, Figures~\ref{fig1}--\ref{fig4}a illustrate the location of the O5--O6 star, 5 GHz radio peaks, 6.7-GHz maser, 
ionized emission, molecular emission, filaments, IRDC, and dust (warm and cold) emission in the complex. 
\subsubsection{H$_{2}$ emission}
In Figure~\ref{fig4}b, we display a continuum-subtracted 2.12 $\mu$m H$_{2}$ (v = $1-0$ S(1)) image near the waist of the nebula 
(size of the selected region $\sim$5.7 pc $\times$ 5.4 pc).
The map shows H$_{2}$ emission toward the waist and edges of the bipolar nebula. 
The comparison of H$_{2}$ emission with the 8.0 $\mu$m features can be seen in Figures~\ref{fig4}b and~\ref{fig5}.  
The 8.0 $\mu$m emission, which contains 7.7 $\mu$m and 8.6 $\mu$m polycyclic aromatic hydrocarbon (PAH) features including the continuum, traces a 
photodissociation region (PDR). Considering the distribution of H$_{2}$ emission, the ionized gas (see Figure~\ref{fig1}) and the 8.0 $\mu$m emission, 
we suggest that the H$_{2}$ emission likely traces the PDR in the complex. 
However, it is also possible that the H$_{2}$ emission probably originates in the shock due to the expanding H\,{\sc ii} region (see Section~\ref{sec:coem}). 
The origin of the H$_{2}$ emission in terms of shocks and UV fluorescence is often explained in the literature by the observed ratio of the 
1--0S(1) to the 2--1S(1) intensity. However, we do not have observations of the source in the 2--1S(1) line of H$_{2}$ filter. 
Therefore, we cannot conclude the origin of H$_{2}$ emission in this work.
Additionally, the H$_{2}$ emission near the 6.7-GHz maser emission at a parsec scale cannot be ruled out 
due to outflow activity \citep[see][and references therein]{dewangan15b}. 
\citet{lee14} and \citet{shinn14} reported the detection of [Fe~II] emission in the southwest of the 6.7-GHz maser emission, 
which further suggests the presence of shocks \citep[see Figure~12 in][]{lee14}. 
Note that the H$_{2}$ emission is also detected at the tip of the southern condensation (see arrow in Figure~\ref{fig4}b). The 8.0 $\mu$m and the H$_{2}$ emissions have very similar spatial structure at the tip of the southern condensation.
 
In summary, the presence of H$_{2}$ features in the complex provides the observational signatures of the outflow activity as well as the impact of the UV photons.
\subsection{Temperature and column density maps of W42 complex}
\label{subsec:temp}
In previous sections, we qualitatively studied the morphology of the complex using {\it Herschel} images. 
In this section we present the temperature and column density maps of the complex, derived using {\it Herschel} images. 
A knowledge of the temperature distribution is important to infer the physical conditions in the cloud. 
The distribution of column density allows to infer the extinction, mass, and density variations in the cloud. 
The final temperature and column density maps were obtained following the same procedures as described in \citet{mallick15}.
These maps were determined from a  pixel-by-pixel spectral energy distribution (SED) fit with a modified blackbody curve to the cold 
dust emission in the {\it Herschel} 160--500 $\mu$m wavelengths regime. We did not include {\it Herschel} 70 $\mu$m data, because 
the 70 $\mu$m emission comes from UV-heated warm dust.

Here, we give a brief description of the procedures. 
In the first step, we converted the surface brightness unit of 250--500 $\mu$m images to Jy pixel$^{-1}$, same as the unit of 160 $\mu$m image.
Next, the final processed 160, 250, and 350 $\mu$m images were convolved to the angular resolution of 
the 500 $\mu$m image ($\sim$37$\arcsec$), 
using the convolution kernels available in HIPE software and then regridded on a 14$\arcsec$ raster. 
We then estimated a background flux level. 
The dark and featureless area far from the main cloud complex is selected for the background estimation. 
The background flux level was obtained to be -2.98, 1.14, 0.55, and 0.19 Jy pixel$^{-1}$ for the 160, 250, 350, and 
500 $\mu$m images (size of the selected region $\sim$6$\farcm$6 $\times$ 7$\farcm$3; 
central coordinates: $\alpha_{J2000}$ = 18$^{h}$42$^{m}$34$^{s}$.6, 
$\delta_{J2000}$ = -06$\degr$23$\arcmin$11$\arcsec$.4), respectively. 
Finally, a modified blackbody was fitted to the observed fluxes on a pixel-by-pixel basis to generate 
the maps \citep[see equations 8 and 9 given in][]{mallick15}. The fitting was done using the four data points for each pixel, keeping the 
dust temperature (T$_{d}$) and the column density ($N(\mathrm H_2)$) 
as free parameters. 
In the calculations, we adopted the mean molecular weight per hydrogen molecule ($\mu_{H2}$=) 2.8 
\citep{kauffmann08} and an absorption coefficient ($\kappa_\nu$ =) 0.1~$(\nu/1000~{\rm GHz})^{\beta}$ cm$^{2}$ g$^{-1}$, 
including a gas-to-dust ratio ($R_t$ =) of 100, with a dust spectral index of $\beta$\,=\,2 \citep[see][]{hildebrand83}.

\begin{figure*}
\epsscale{1.0}
\plotone{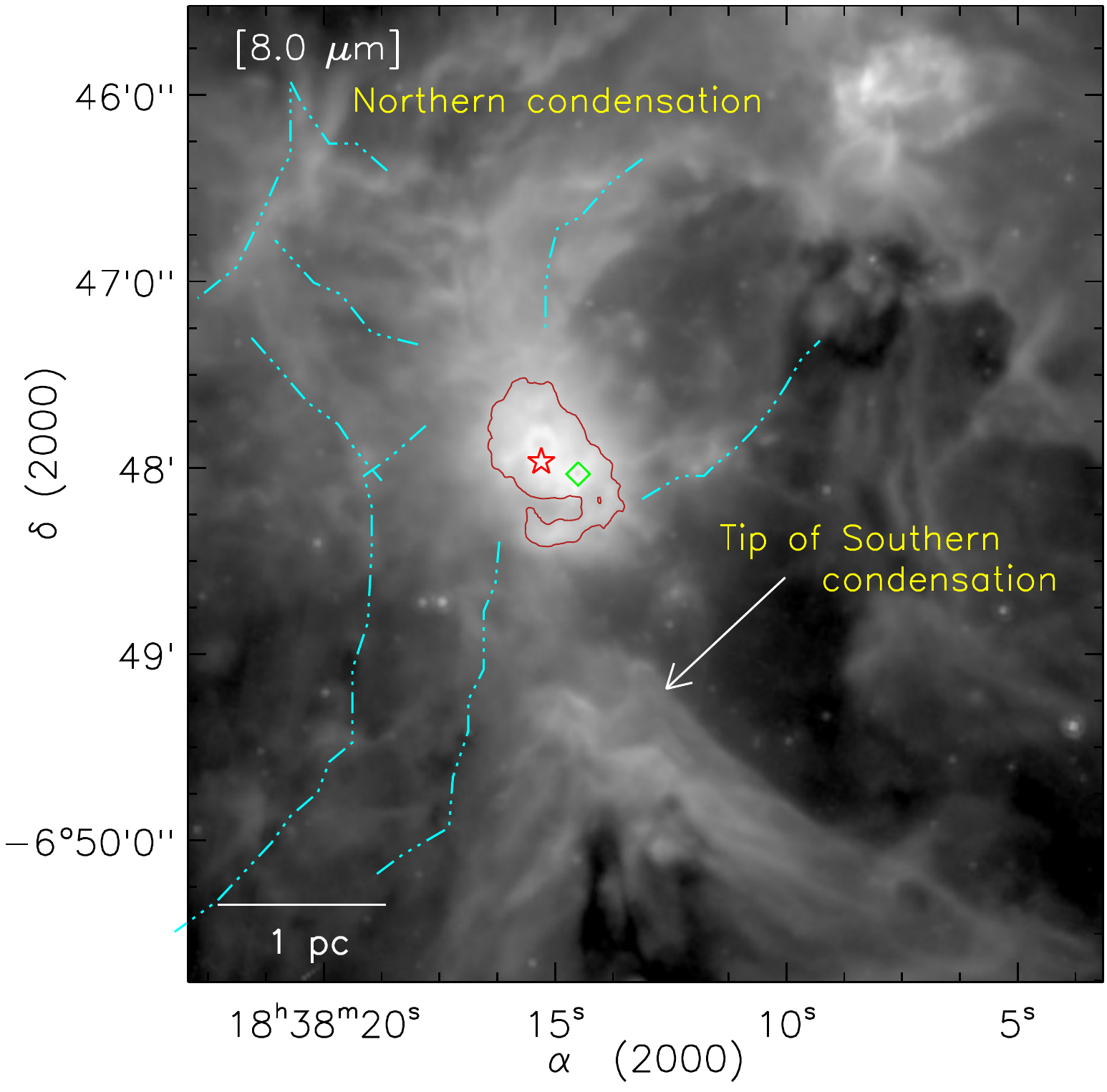}
\caption{\scriptsize {\it Spitzer} 8.0 $\mu$m continuum image in log scale, as shown by a magenta box in Figure~\ref{fig3}f. 
The area shown in H$_{2}$ and 8.0 $\mu$m images is similar. 
{\it Spitzer} 5.8 $\mu$m contour emission (in firebrick color) is shown with a level of 555 MJy/Sr. 
Filaments as seen in the {\it Herschel} maps (Figure~\ref{fig4}a) are also overplotted by dot-dashed cyan color curves. 
The other marked symbols are similar to those shown in Figure~\ref{fig4}b.}
\label{fig5}
\end{figure*}

\begin{figure*}
\epsscale{0.73}
\plotone{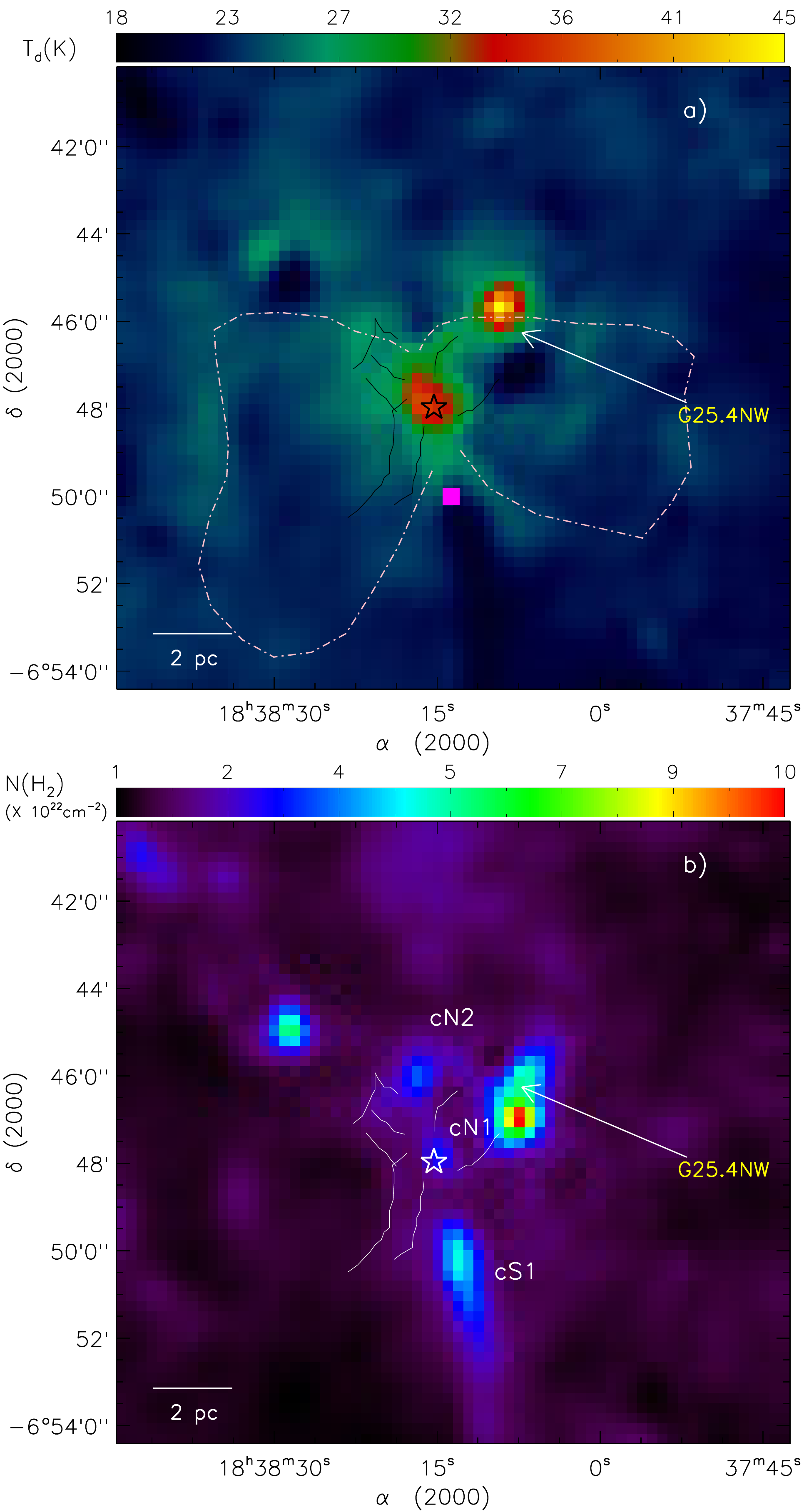}
\caption{\scriptsize a) {\it Herschel} temperature map of W42 complex (see text for details). 
A bipolar nebula is also highlighted similar to the one shown in Figure~\ref{fig1}. 
A filled magenta square refers an ATLASGAL clump, which has NH3 line parameters from the NH3 survey of dense ATLASGAL clumps 
\citep[see][]{wienen12}.
b) {\it Herschel} column density ($N(\mathrm H_2)$) map of W42 complex (see text for details). The map also provides the information of extinction with 
$A_V=1.07 \times 10^{-21}~N(\mathrm H_2)$. Two clumps in the northern condensation (cN1 and cN2) and one clump (cS1) in the southern 
condensation are marked in the figure. In both the panels, the star symbol indicates the location of an O5-O6 star. 
The {\it Herschel} filaments are also overplotted by solid curves in both the panels (see Figure~\ref{fig4}a).}
\label{fig6}
\end{figure*}

The final temperature and column density maps (angular resolution $\sim$37$\arcsec$) are shown in Figures~\ref{fig6}a and~\ref{fig6}b, respectively. 
The temperature map clearly shows temperature variations in the complex. 
Our temperature map shows considerably warmer gas (T$_{d}$ $\sim$30-36 K) towards the heart of the bipolar nebula, 
where the ionizing source is located. 
We find that the mean temperature of most of the nebula is $\sim$25~K in the complex.  
\citet{wienen12} derived the line parameters (such as the gas kinetic temperature and rotational temperature) 
towards the clumps found by the ATLASGAL survey. We found one ATLASGAL dust 
clump in our selected region (see Figure~\ref{fig6}a), 
which has NH$_{3}$ line parameters from \citet{wienen12}. The rotational temperature of this clump was found to be $\sim$20 K. 
It is consistent with our estimated temperature value towards this clump. 
G25.4NW region also dominates in the temperature map with a peak temperature of $\sim$45 K. 
Two clumps, as marked in the CO map (cN1 and cN2; see Figure~\ref{fig3}f), are traced in the northern condensation with peak column densities of $\sim$3.6~$\times$~10$^{22}$ and 3.0~$\times$~10$^{22}$ cm$^{-2}$, which correspond to visual extinctions of A$_{V}$ $\sim$38.5--32 mag, assuming 
the classical relation from \citet{bohlin78} (i.e. $A_V=1.07 \times 10^{-21}~N(\mathrm H_2)$).
The colder gas (T$_{d}$ $\sim$20 K) is found near the column density peak (5~$\times$~10$^{22}$ cm$^{-2}$; A$_{V}$ $\sim$53.5 mag) in the southern condensation clump (cS1), where the IRDC is seen in 8.0 $\mu$m map (see Figure~\ref{fig6}).  
\citet{anderson09} studied the molecular cloud U25.38$-$0.18 (i.e. northern condensation) using the GRS CO data and estimated its column density to be $\sim$1.05 $\times$ 10$^{22}$ cm$^{-2}$, which is well in agreement with our column density 
value towards the northern condensation. 
The condensation associated with G25.4NW region has the highest column density, as traced in Figure~\ref{fig6}b.  
{\it Herschel} filaments are also drawn in Figure~\ref{fig6}. 
Due to the coarse resolution of the column density map, some of the filaments are not resolved as seen in the 250 $\mu$m map. 
The value of column densities towards these filaments is found to be $\sim$1.5~$\times$~10$^{22}$ cm$^{-2}$. 
The column density map shows the dense clump (cN1; peak $N(\mathrm H_2)$ $\sim$3.0~$\times$~10$^{22}$ cm$^{-2}$) associated 
with the O5--O6 star and W42-MME, where several filaments (``a--h"; see Figure~\ref{fig4}a) appear to be radially 
directed to this clump (see Figure~\ref{fig6}b and also Figure~\ref{fig5}), revealing a ``hub-filament" morphology \citep[e.g.][]{myers09}. 
We should mention here that all these filaments are located well within the W42 molecular cloud, except the two filaments ``h" and ``g" 
that appear to be extended towards G25.4NW from the northern condensations (also see Figure~\ref{fig5}). 
We suggest that this could be a projection effect because these two 
regions (W42 and G25.4NW) are not physically linked. Four filaments (e, f, g, and h) are also seen in the H$_{2}$ and 8.0 $\mu$m maps, 
while remaining filaments (a, b, c, and d) are traced only in the {\it Herschel} images. 
These filaments are neither perfectly parallel nor perpendicular to the parental molecular cloud. 
The cavity-like structure (embedded in cN1 clump) appears at the junction of filaments (see Figure~\ref{fig5}). 
This cavity seems to be an important link for understanding the ongoing physical processes in the complex.

Using the column density map (in Figure~\ref{fig6}b), we also estimated the total masses of the clumps associated with the northern 
condensation (cN1 and cN2) and the southern condensation (cS1). 
Total column densities ($\Sigma N(\mathrm H_2)$) of these clumps were computed using the ``CLUMPFIND" 
IDL program \citep{williams94}. 
Following the relation M $\propto$ $\Sigma N(\mathrm H_2)$ \citep[also see][]{mallick15}, the masses of the clumps cN1, cN2, and cS1 are 
computed to be $\sim$730, $\sim$840, and $\sim$2564 $M_\odot$, respectively. 
\subsection{$^{13}$CO (J=1$-$0) kinematics in W42 complex}
\label{sec:coem} 
\begin{figure*}
\epsscale{1.0}
\plotone{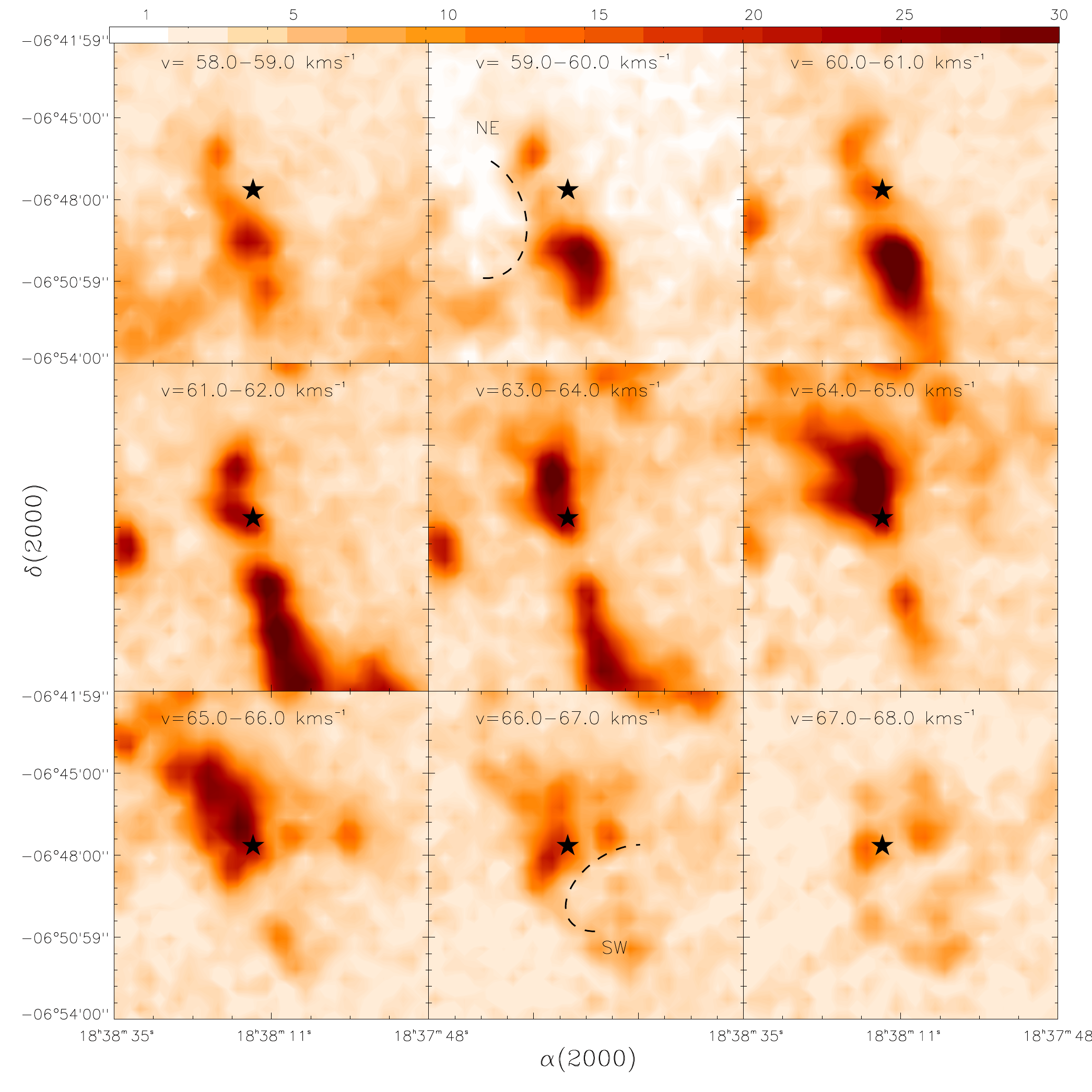}
\caption{\scriptsize The $^{13}$CO (J=1$-$0) velocity channel maps of W42 complex. 
Each channel map is obtained by integrating five original channels between the 
velocity coverage range indicated in each panel. 
The color bar is shown from $1\sigma$ at intervals of $1\sigma$, where the channel rms is 0.06 K km s$^{-1}$. 
The position of an O5-O6 star is marked with a filled star symbol. 
Two dashed curves (in black color) represent the regions empty of molecular gas (also see Figure~\ref{fig9}).} 
\label{fig7} 
\end{figure*}

The $^{13}$CO profile showed that the W42 complex is covered in the velocity range of 58--69 km s$^{-1}$. 
In Figure~\ref{fig7}, we present the integrated GRS $^{13}$CO (J=1$-$0) velocity channel maps (at intervals of 1 km s$^{-1}$), 
which reveal the morphology and different molecular components along the line of sight. 
The velocity channel maps trace the prominent northern and southern condensations in the complex. 
The ionizing source is located within the northern condensation. The physical association of the ionizing source 
was confirmed by the velocities of molecular and ionized gas, as mentioned in the introduction.
Additionally, the maps reveal the regions empty of molecular CO gas (see highlighted regions in Figure~\ref{fig7}).

\begin{figure*}
\epsscale{1.0}
\plotone{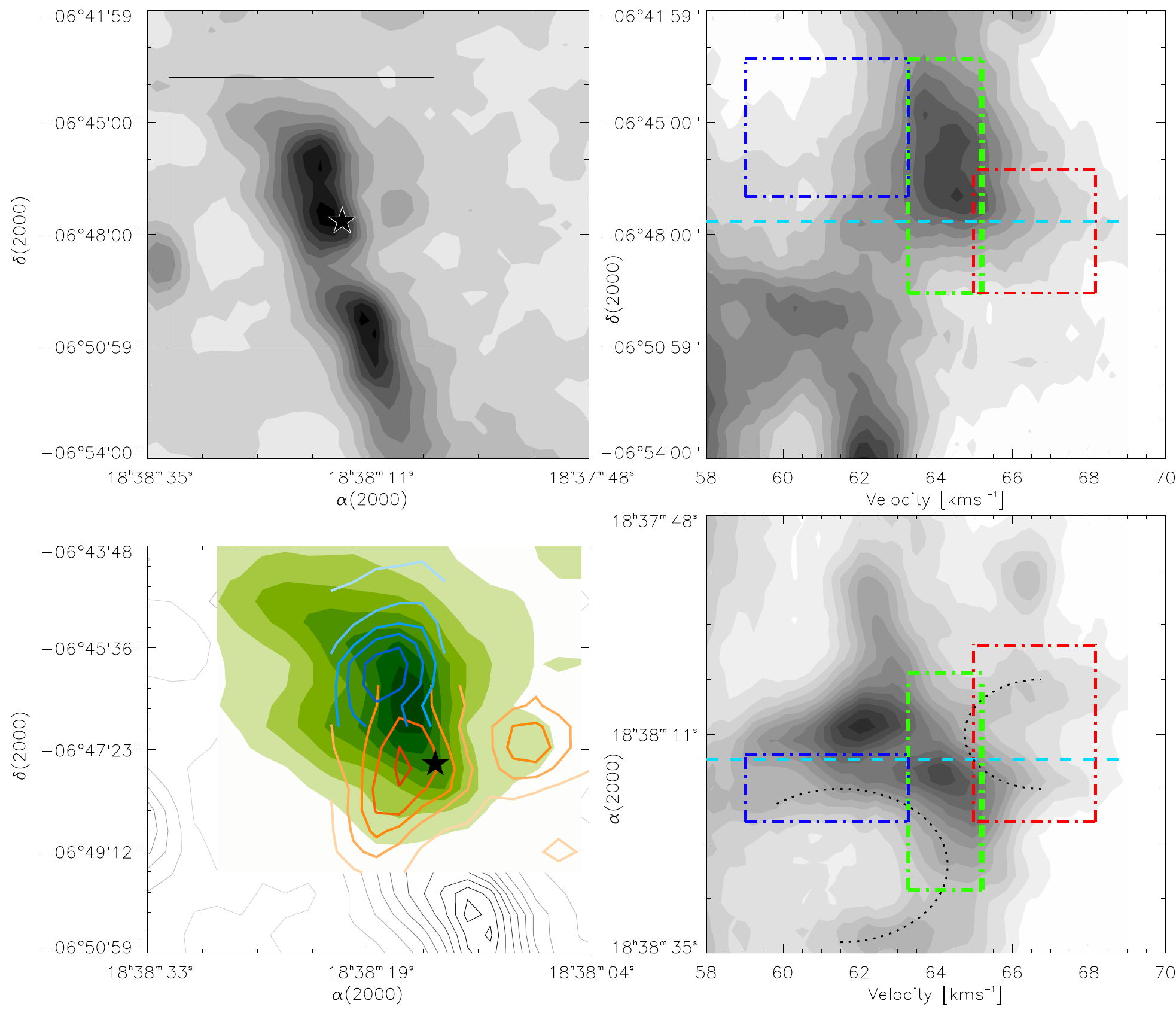}
\caption{\scriptsize Two-dimensional projections of the GRS $^{13}$CO (J=1$-$0) data cube. 
{\bf Top Left:} Integrated intensity map of W42 complex. 
The molecular emissions are integrated over the velocity range from 58 to 69 km s$^{-1}$. 
The lowest gray-scale level corresponds to $20\sigma$, with successive levels 
increasing in steps of $20\sigma$. The channel rms is 0.201 K km s$^{-1}$. 
A zoomed-in view of the solid box is shown in bottom left panel. 
The position of an O5-O6 star is marked with a star symbol. 
{\bf Bottom Left:} Contour maps show the receding gas (65--68~km\,s$^{-1}$; red) 
and approaching gas (59--63~km\,s$^{-1}$; blue) with respect to that at 
rest (63--65~km\,s$^{-1}$; green color) for the highlighted region in top left panel. 
The lowest level corresponds to $20\sigma$ (black; $1\sigma$ = 0.201~K km\,s$^{-1}$), $25\sigma$ (red; $1\sigma$ = 0.107~K km\,s$^{-1}$), 
 $30\sigma$ (green; $1\sigma$ = 0.082~K km\,s$^{-1}$), and $25\sigma$ (blue; $1\sigma$ = 0.123~K km\,s$^{-1}$), 
 with successive levels increasing in steps of $20\sigma$ (black), $25\sigma$ (red), $30\sigma$ (green), and $25\sigma$ (blue). 
 The position of an O5-O6 star is shown with a filled star symbol. 
 {\bf Top Right:} Declination-velocity map. The CO emission is integrated over the right ascension range from 
 18$^{h}$37$^{m}$48$^{s}$ (279.45$\degr$) to 18$^{h}$38$^{m}$35$^{s}$ (279.65$\degr$). 
 {\bf Bottom Right:} Right Ascension-velocity map. The CO emission is integrated over the declination range from  $-$06$\degr$54$\arcmin$00$\arcsec$ ($-$6.9$\degr$) to $-$06$\degr$42$\arcmin$00$\arcsec$ ($-$6.7$\degr$). 
 The semi-ring-like or inverted C-like features are highlighted by dotted curves (in black color) (also see the text).
 In position-velocity maps (both right panels), the lowest gray-scale level corresponds to $10\sigma$ 
 (where $1\sigma$ = 0.005 K degree), with successive levels increasing in steps of $10\sigma$. 
The slices of receding, approaching, and rest gas are highlighted in red, blue, and green dot-dashed boxes, respectively. 
Cyan color dashed line shows the position of the ionizing star in the position-velocity maps.} 
\label{fig8} 
\end{figure*}

The integrated $^{13}$CO intensity map and position-velocity maps are shown in Figure~\ref{fig8}.
The position-velocity maps (right ascension-velocity and declination-velocity) show the presence of 
a noticeable velocity gradient in both the northern and southern molecular components. 
The position-velocity plots of $^{13}$CO gas reveal an almost semi-ring-like or inverted C-like structure 
(see Right Ascension-velocity panel in Figure~\ref{fig8}).  
Such a structure in the position-velocity plot is consistent with the model for an expanding shell \citep{arce11}. 
\citet{arce11} performed modeling of expanding bubbles in a turbulent medium and compared with the observed 
structures in Perseus molecular cloud. They suggested that the semi-ring-like or C-like structure is characteristic of an expanding shell.
These authors also pointed out that a ring-like structure can be evident in the position-velocity plot 
when the powering source is located at the center of the region. Since the complex harbors a powerful O5--O6 star. 
Therefore, the existence of an inverted C-like structure can be explained by the expanding H\,{\sc ii} region. 
Following the position-velocity maps, we infer the expansion velocity of the gas to be $\sim$3 km s$^{-1}$. 

We further analyzed the gas distribution in the northern component using the position-velocity analysis and found 
the receding gas (65--68~km\,s$^{-1}$), approaching gas (59--63~km\,s$^{-1}$), and rest gas (63--65~km\,s$^{-1}$) components (see bottom left panel in Figure~\ref{fig8}). 
Due to the coarse beam of CO data (beam size $\sim$45$\arcsec$), we cannot pinpoint the exact exciting source of this outflow. 
This outflow signature is associated with whole northern condensation, hence it could be related to the expanding H\,{\sc ii} region associated with the O5--O6 star.

\begin{figure*}
\epsscale{1.0}
\plotone{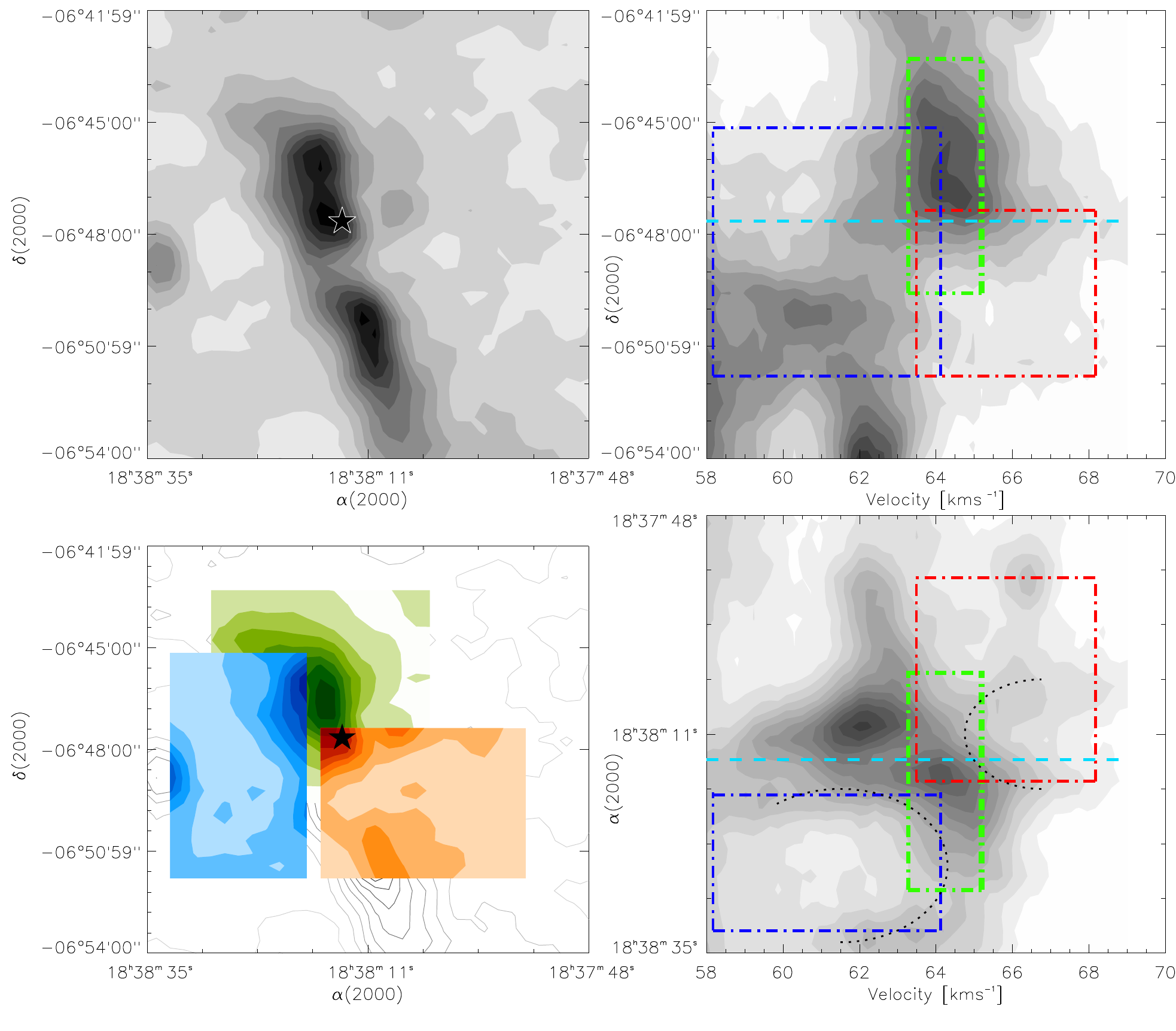}
\caption{\scriptsize Two-dimensional projections of the GRS $^{13}$CO (J=1$-$0) data cube similar to those shown in figure~\ref{fig8}. 
{\bf Top Left:} Integrated intensity map of W42 complex. 
The position of an O5-O6 star is marked with a star symbol. 
{\bf Bottom Left:} Contour maps show the receding gas (63.5--68.1~km\,s$^{-1}$; red) 
and approaching gas (58.2--64.1~km\,s$^{-1}$; blue) with respect to that at 
rest (63.3--65.2~km\,s$^{-1}$; green color). The empty molecular gas regions (i.e. cavities) are observed (also see figure~\ref{fig9}). 
 The position of an O5-O6 star is shown with a filled star symbol. 
{\bf Right panels:} the slices of receding, approaching, and rest gas are highlighted in red, blue, and green dot-dashed boxes, respectively. 
Cyan color dashed line shows the position of the ionizing star in the position-velocity maps.} 
\label{fig8a} 
\end{figure*}

In order to trace out the regions of direct interaction of ionized gas, we carefully analyzed the position-velocity maps 
(see Figure~\ref{fig8a}) and 
found two cavities empty of molecular gas in the south-west and south-east directions with respect to 
the O5-O6 star (see Figures~\ref{fig8a} and~\ref{fig9}). 
These cavities are traced in the velocity ranges of 63--68~km\,s$^{-1}$ and 58--64~km\,s$^{-1}$ in the south-west and south-east directions, respectively. 
The spatial locations of the cavities show the symmetry with respect to the main elongated molecular cloud.
The existence of the molecular cavities suggests that the molecular gas has been eroded by the ionizing gas.

Note that the GRS $^{13}$CO data do not allow us to explore any outflow signatures towards W42-MME 
due to the coarse beam (beam size $\sim$45$\arcsec$). 
Hence, high resolution molecular line observations are required to obtain better insight into the molecular outflows in the complex.
Combining the inferences from the CO kinematics, the influence of ionizing gas on the surroundings is evident. 
\subsection{Dynamical age of the H\,{\sc ii} region and feedback of a massive star}
\label{sec:dynage} 
In this section, we compute the dynamical age of the H\,{\sc ii} region. 
Knowledge of this physical parameter helps in understanding the local star formation process by the interaction of the H\,{\sc ii} region  with the surrounding ISM. 
Here, we performed the calculations based on the radio continuum flux at 20 cm and radio spectral type of the ionizing source.
The dynamical age (t$_{dyn}$) of a spherically expanding H\,{\sc ii} region of a radius R is estimated 
using the model described by \citet{dyson80} and is given by:
\begin{equation}
t_{dyn} = \left(\frac{4\,R_{s}}{7\,c_{s}}\right) \,\left[\left(\frac{R}{R_{s}}\right)^{7/4}- 1\right] 
\end{equation}
\noindent where c$_{s}$ is the isothermal sound velocity in the ionized gas (c$_{s}$ = 10 km s$^{-1}$) 
and R$_{s}$ (= 3N$_{uv}$/4$\pi n^2_{\rm{0}} \alpha_{B}$)$^{1/3}$ is the radius of the initial Str\"{o}mgren sphere, 
where the initial particle number density of the ambient neutral gas is ``n$_{0}$'', the radiative recombination 
coefficient is ``$\alpha_{B}$'' \citep[=  2.6 $\times$ 10$^{-13}$ $\times$ (10$^{4}$ K/T$_{e}$)$^{0.7}$ cm$^{3}$ s$^{-1}$; see][]{kwan97}, 
and N$_{uv}$ is the Lyman continuum photons per second. 
In this calculation, we assume that the H\,{\sc ii} region associated with the complex was spherical in morphology 
during its initial phase and, with time, it evolved into the homogeneous surrounding environments. 
Here, we use N$_{uv}$ = 1.83 $\times$ 10$^{49}$ s$^{-1}$ or logN$_{uv}$ = 49.26 (see Section~\ref{subsec:data1}) 
for a radius of the H\,{\sc ii} region, R $\approx$ 2.14 pc \citep{beaumont10}, 
and $\alpha_{B}$ = 2.6 $\times$ 10$^{-13}$ cm$^{3}$ s$^{-1}$ at T$_{e}$ = 10000 K.  
\citet{anderson09} estimated the column density (N(H$_{2}$)) and size ($L$) of the 
molecular cloud U25.38$-$0.18 (i.e. northern condensation) as $\sim$1.05 $\times$ 10$^{22}$ cm$^{-2}$ 
and 1$\farcm$5 (5.12 $\times$ 10$^{18}$ cm at a distance of 3.8 kpc), respectively.
The H$_{2}$ number density is computed to be $\sim$2051 cm$^{-3}$ using the relation $N_{\rm H_{2}}$ (cm$^{-2}$)/$L$ (cm), for an assumed spherical 
structure. Using the values of N$_{uv}$, n$_{0}$, and R in Equation~1, 
we obtain R$_{s}$ and dynamical age of the H\,{\sc ii} region as $\sim$0.51 pc and $\sim$0.32 Myr, respectively. 
If we also adopt a distance of 2.2 kpc to W42 
then we obtain a radius of the H\,{\sc ii} region (R) $\approx$ 1.24 pc, N$_{uv}$ = 6.13 $\times$ 10$^{48}$ s$^{-1}$ or logN$_{uv}$ = 
48.79 (see Section~\ref{subsec:data1}), size of the molecular cloud ($L$) $\sim$2.97 $\times$ 10$^{18}$ cm, 
H$_{2}$ number density $\sim$3541 cm$^{-3}$, R$_{s}$ $\sim$0.35 pc, and dynamical age of the H\,{\sc ii} region $\sim$0.22 Myr. 
The dynamical age of the H\,{\sc ii} region calculated at 2.2 kpc is estimated to be about 69\% of that derived at a distance of 3.8 kpc.    
The estimated dynamical age should be considered with some caution, because of the assumptions involving 
spherical geometry and uniform density distribution. Also, the observed 
bipolar morphology of the complex could have originated due to the evolution of the H\,{\sc ii} region in a medium with strong density gradients.

In the previous sections, we found that the massive star (O5--O6) is located within the northern condensation. 
In order to study the feedback of this massive star on the southern condensation, we calculated the following pressure components \citep[e.g.][]{bressert12}:\\
(i) pressure of an H\,{\sc ii} region $(P_{HII}) = \mu m_{H} c_{s}^2\, \left(\sqrt{3N_{uv}\over 4\pi\,\alpha_{B}\, D_{s}^3}\right)$;\\ 
(ii) radiation pressure (P$_{rad}$) = $L_{bol}/ 4\pi c D_{s}^2$; \\ 
(iii) stellar wind ram pressure (P$_{wind}$) = $\dot{M}_{w} V_{w} / 4 \pi D_{s}^2$; \\
(iv) pressure exerted by the self-gravity of the surrounding molecular gas $(P_{scloud}) \approx\pi G\Sigma^2$ \citep[e.g.][]{harper09}.\\
In the relations above, N$_{uv}$ and $\alpha_{B}$ are defined as in Equation~1, 
$\mu$ = 2.37 (approximately 70\% H and 28\% He by mass), m$_{H}$ is the hydrogen atom mass, c$_{s}$ is the sound 
speed of the photo-ionized gas (= 10 km s$^{-1}$), $\dot{M}_{w}$ is the mass-loss rate, 
V$_{w}$ is the wind velocity of the ionizing source, L$_{bol}$ is the bolometric luminosity of the region, 
$\Sigma$ $(= M_{scloud}/\pi R_{c}^2)$ is the mean mass surface density of the southern condensation, 
M$_{scloud}$ is the mass of the molecular gas associated with the southern condensation, and R$_{c}$ is the radius of the molecular region. 
The pressure components associated with massive star are evaluated at D$_{s}$ = 2.14 pc (radius of the H\,{\sc ii} region) on the southern condensation 
from the position of the O5-O6 type star. 
The luminosity of the exciting O5 type star is $\sim$3.2 $\times$ 10$^{5}$ L$_{\odot}$ 
\citep[logL/L$_{\odot}$ = 5.51; see Table~1 given in][]{martins05}. 

Substituting $M_{scloud}$ (clump cS1) $\approx$ 2564 $M_\odot$ (see Section~\ref{subsec:temp}), R$_{c}$ $\approx$ 1.6 pc, 
$L_{bol}$ = 3.2 $\times$ 10$^{5}$ L$_{\odot}$, 
$\dot{M}_{w}$ = 2.0 $\times$ 10$^{-7}$ M$_{\odot}$ yr$^{-1}$ \citep[for an O6V star;][]{dejager88}, 
V$_{w}$ = 2500 km s$^{-1}$ \citep[for an O6V star;][]{prinja90} in the above equations, 
we find a surface density $\Sigma \approx$ 0.067 g cm$^{-2}$, 
$P_{scloud}$ $\approx$ 9.3 $\times$ 10$^{-10}$ dynes cm$^{-2}$, 
P$_{HII}$ $\approx$ 9.3 $\times$ 10$^{-10}$ dynes\, cm$^{-2}$, 
$P_{rad}$ $\approx$ 7.57 $\times$ 10$^{-11}$ dynes\, cm$^{-2}$, and
P$_{wind}$ $\approx$ 5.76 $\times$ 10$^{-12}$  dynes\, cm$^{-2}$. 
The comparison of different pressure components associated with the massive star suggests that the pressure 
of the H\,{\sc ii} region is relatively higher than the radiation pressure and the stellar wind pressure. 

Note that a typical cool molecular cloud (temperature $\sim$20 K and particle density $\sim$10$^{3}$--10$^{4}$ cm$^{-3}$) 
has pressure values $\sim$10$^{-11}$--10$^{-12}$ dynes cm$^{-2}$ \citep[see Table 7.3 of][]{dyson80}. 
We estimate the value of $P_{scloud}$ $\approx$ 9.3 $\times$ 10$^{-10}$ dynes cm$^{-2}$, which is relatively higher than 
the pressure associated with a typical cool molecular cloud.  
It suggests that the surrounding molecular cloud has been compressed to increase the pressure. 

Furthermore, we also derived the virial mass of the southern condensation using the line width ($\Delta$v) of $^{13}$CO velocity profile 
obtained using a single Gaussian fitting. The virial mass is given in \citet{maclaren88} as M$_{vir}$ ($M_\odot$)\,=\,k\,R$_{c}$\,$\Delta$v$^2$, 
with R$_{c}$ (in pc) is the radius of the clump as defined above, $\Delta$v (in km s$^{-1}$) is the line width = 2.8 km s$^{-1}$, 
and the geometrical parameter, k\,=\,126, for a density profile $\rho$ $\propto$ 1/r$^2$. 
We therefore find M$_{vir}$\,= 1580 $M_\odot$ for k\,=\,126. 
In general, the clump can be stable against gravitational collapse if the virial ratio $M_{scloud}$/M$_{vir}$~$\sim$~1.
The virial ratio conditions, $M_{scloud}$/M$_{vir}$~$<$ 1 and $M_{scloud}$/M$_{vir}$~$>$ 1 
provide the signatures of unbound clump and unstable clump against gravity, respectively. 
In the present case, the ratio $M_{scloud}$/M$_{vir}$ appears larger than unity, which indicates unstable clump against gravitational collapse. 
Star formation activity is traced in this clump (see Section~\ref{subsec:surfden} for more details), 
hence the higher $P_{scloud}$ value could be explained due to the effects of self-gravity.

Note that the dynamical age of the H\,{\sc ii} region is $\sim$0.32 Myr. 
Therefore, the W42 H\,{\sc ii} region might not have influenced the vicinity prior to this time period.
The total pressure (P$_{total}$ = P$_{HII}$ + $P_{rad}$ + P$_{wind}$) driven by a massive star is found to be 
$\sim$1.0 $\times$ 10$^{-9}$ dynes\, cm$^{-2}$, 
which is comparable with the pressure exerted by the surrounding, self-gravitating molecular cloud ($\approx$ 9.3 $\times$ 10$^{-10}$ dynes cm$^{-2}$).
This particular result provides the evidence that the southern condensation is not 
destroyed by the impact of the expanding H\,{\sc ii} region. 
These conclusions are valid even for the distance of 2.2 kpc to W42. If we compute different pressures for this distance then we 
obtain P$_{HII}$ = 2.1 $\times$ 10$^{-9}$ dynes\, cm$^{-2}$, $P_{rad}$ = 2.3 $\times$ 10$^{-10}$ dynes\, cm$^{-2}$, 
P$_{wind}$ = 1.7 $\times$ 10$^{-11}$ dynes\, cm$^{-2}$, P$_{total}$ = 2.3 $\times$ 10$^{-9}$ dynes\, cm$^{-2}$, and 
$P_{scloud}$ = 7.8 $\times$ 10$^{-9}$ dynes\, cm$^{-2}$.   
\subsection{Infrared excess populations}
\subsubsection{Identification and classification of infrared excess sources}
\label{subsec:phot1}
YSOs are identified based on their infrared excess emission. 
Here, we summarize the different schemes to identify and classify YSOs using MIPSGAL, IRAC, and WFCAM photometric data.\\\\
$\bullet$ {\it IRAC-MIPSGAL bands:} 
\citet{guieu10}, \citet{rebull11}, and \citet{samal15} utilized a [3.6] $-$ [24]/[3.6] color-magnitude plot to 
identify YSOs using IRAC and MIPSGAL bands (i.e. IRAC-MIPSGAL). 
Firstly, we identified MIPSGAL 24 $\mu$m sources that are common with IRAC point sources. 
Note that since MIPSGAL 24 $\mu$m image is saturated near the IRAS location, this scheme 
cannot provide information of YSOs towards the main central ionizing area. 58 sources are found to be common in the IRAC-MIPSGAL bands. 
The IRAC-MIPSGAL color-magnitude diagram ([3.6] $-$ [24]/[3.6]) is shown in Figure~\ref{fig10a} for all the identified sources. 
Using this scheme, we find 27 candidate YSOs (7 Class I; 3 Flat-spectrum; 17 Class~II) and 31 candidate Class~III sources in our selected region. 
In Figure~\ref{fig10a}, we marked different zones occupied by YSOs and Class~III sources \citep[also see Figure~6 in][]{guieu10}.   
Figure also shows the extinction vector with A$_{K}$ = 5 mag which is obtained using the average extinction laws 
(A$_{3.6\mu m}$/A$_{K}$ = 0.632 and A$_{24\mu m}$/A$_{K}$ = 0.48) from \citet{flaherty07}. 
Note that if some Class~II YSOs suffer extinction (A$_{V}$) of about more than 20 mag then such sources could be shifted in the 
location of the Flat-spectrum. 
We also checked our sources for possible contaminants (i.e. galaxies and disk-less stars) using the color-magnitude space of 
the SWIRE field \citep[see Figure~10 in][]{rebull11}. We do not find any contaminants (i.e. galaxies) in our selected sources 
(see Figure~\ref{fig10a}). One can find more details about the zones of contaminants in the color-magnitude space in the work of \citet{rebull11}.\\
$\bullet$ {\it Four IRAC bands:} 
\citet{gutermuth09} developed YSO classification methods using four IRAC bands. 
Using IRAC colors, they also identified various possible contaminants (e.g. broad-line active galactic nuclei (AGNs), 
PAH-emitting galaxies, shocked emission blobs/knots, and PAH-emission-contaminated apertures). 
In order to identify YSOs and likely contaminants, we adopted the various color criteria suggested by \citet{gutermuth09}.
The selected candidate YSOs were further classified into different evolutionary stages (i.e. Class~I, Class~II, and Class~III), 
using the slopes of the IRAC SED ($\alpha_{IRAC}$) measured from 3.6 to 8.0 $\mu$m 
\citep[e.g.,][]{lada06}. The details of YSO classifications can also be found in \citet[][and references therein]{dewangan11}. 
The IRAC color-color diagram ([3.6]$-$[4.5] vs [5.8]$-$[8.0]) is shown in Figure~\ref{fig10b}a for all the identified sources. 
Following this procedure, we find 39 candidate YSOs (14 Class I; 25 Class~II), 1 candidate Class~III, 1285 photospheres, 
and 123 contaminants in the selected region.\\
$\bullet$ {\it WFCAM-IRAC bands:} 
WFCAM-IRAC (H, K, 3.6, and 4.5 $\mu$m) classification method is adopted for the sources, that are 
not detected in two longer wavelengths of IRAC bands (5.8 and 8.0 $\mu$m). One can find more details about this method in \citet{gutermuth09}.
In this method, the dereddened colors ([K$-$[3.6]]$_{0}$ and [[3.6]$-$[4.5]]$_{0}$) were estimated using the color 
excess ratios given in \citet{flaherty07}. 
Additional conditions (i.e., [3.6]$_{0}$ $<$ 14.5 mag for Class~II and [3.6]$_{0}$ $<$ 15 mag for Class~I) were also applied on the 
identified YSOs (Class~I and Class~II) to check for possible dim extragalactic contaminants. 
The dereddened 3.6 $\mu$m magnitudes were obtained using observed color and the extinction law from \citet{flaherty07}. 
We obtain 252 candidate YSOs (16 Class~I and 236 Class~II) using WFCAM-IRAC data (see Figure~\ref{fig10b}b). \\
$\bullet$ {\it Three IRAC bands:}
One can identify additional protostars using only three IRAC bands (3.6, 4.5, and 5.8 $\mu$m), 
when sources are not detected or saturated in 8.0 $\mu$m band. 
Using three IRAC bands, \citet{hartmann05} and \citet{getman07} identified protostars with the criteria [3.6]$-$[4.5] $\ge$ 0.7 and [4.5]$-$[5.8] $\ge$ 0.7. 
Following this approach, we identify 9 candidate protostars (see Figure~\ref{fig10b}c). \\
$\bullet$ {\it H-K color excess:} Those sources detected only in the NIR regime can be used to identify YSOs, having a large color excess in H$-$K.
Red sources (having H$-$K $>$ 2.35) were also identified using the color-magnitude (H$-$K/K) diagram (see Figure~\ref{fig10b}d). 
This color criterion is selected from the color-magnitude analysis of the nearby control field 
(size $\sim$5$\farcm$4  $\times$ 5$\farcm$4; central coordinates: $\alpha_{J2000}$ = 18$^{h}$38$^{m}$41$^{s}$.2, 
$\delta_{J2000}$ = -06$\degr$53$\arcmin$53$\arcsec$.8). We generated a color-magnitude (H$-$K/K) diagram for our selected control field and 
used all sources having detections in H and K bands. We obtained a color H$-$K value (i.e. $\sim$2.35) that 
separates large H$-$K excess sources from the rest of the population.
This cut-off condition leads to 185 additional deeply embedded sources in the complex.

All the four schemes yield a total of 512 candidate YSOs in the complex. 
The positions of all YSOs are shown in Figure~\ref{fig11}.
\subsubsection{Spatial distribution of YSOs}
\label{subsec:surfden}
In this section, we combine all the selected candidate YSOs from different schemes to examine 
their spatial distributions in the complex.
To study the spatial distribution of candidate YSOs, we generate their surface density map.
The surface density map can be constructed by dividing the mosaic image using a regular grid and estimating 
the surface density of candidate YSOs at each point of the grid. The formula of surface number density at the {\it i$^{th}$} grid 
point is given by $\rho_{i} = (n-1)/A_{i}$ \citep[e.g.][]{casertano85}, where $A_{i}$ is the surface area defined by the radial distance to 
the $n$ = 6 nearest neighbor (NN). The map was created using a 5$\arcsec$ grid at a distance of 3.8 kpc, which is shown as contours in Figure~\ref{fig12}.
The surface density contours are drawn at levels of 3$\sigma$ (4 YSOs/pc$^{2}$; where 1$\sigma$ = 1.4 YSOs/pc$^{2}$), 4$\sigma$ (6 YSOs/pc$^{2}$), 6$\sigma$ (8 YSOs/pc$^{2}$), and 9$\sigma$ (13 YSOs/pc$^{2}$), 
increasing from the outer to the inner regions. 
More details on the surface density of YSOs can be found in the work of \citet{gutermuth09} and \citet{dewangan11}. 
Figure~\ref{fig12} shows the spatial correlation between YSO surface density, molecular gas, and filaments. 
Note that if we use a distance of 2.2 kpc to W42, then we obtain the same surface density structure using the same contour levels 
(i.e. 3$\sigma$, 4$\sigma$, 6$\sigma$, and 9$\sigma$). 
However, the only difference is the value of 1$\sigma$ = 4.2 YSOs/pc$^{2}$ at a distance of 2.2 kpc.

To estimate the clustered YSO populations, we employed an empirical cumulative distribution (ECD) of YSOs as a 
function of NN distance. We select a cutoff length (also referred as the distance of inflection d$_{c}$) using the ECD, 
which allows to delineate the low-density populations \citep[see][for more details]{chavarria08,gutermuth09,dewangan11,dewangan15}.
The analysis yielded a cutoff distance of d$_{c}$ $\sim$0$^{\degr}$.01262 (or $\sim$0.862 pc) 
to identify the cluster members within the contour level of 3$\sigma$ (4 YSOs/pc$^{2}$) in the entire region. 
A cutoff distance results in a clustered fraction of $\sim$40\% YSOs (i.e. 206 from a total of 512 YSOs). 
The GRS $^{13}$CO data allowed us to infer the exact boundary of the W42 molecular cloud. In Figure~\ref{fig12}, 
YSO clusters (i.e. g1, g2, g4, and g5) are spatially distributed well within the W42 molecular cloud, 
which confirms their association with W42. However, the YSO cluster g3 lies close to G25.4NW. 
In this direction several molecular clouds are overlapping, and the molecular clump close to G25.4NW 
has the highest column density of the field. Therefore, some of the YSOs in this direction may not be part of the cluster g3 and they 
could be associated with the clump close to G25.4NW. 
Additionally, the remaining YSO clusters located away from the W42 molecular cloud could be situated at larger distances. 
In general, the study of the distribution of molecular gas and YSOs is considered as a useful tool to overcome the projection effect. 
The positions of Class~I candidate YSOs are also shown in Figure~\ref{fig12}. 
Additionally, two embedded sources (positions and 24 $\mu$m photometry: 1. $\alpha_{2000}$ =18:38:12.7, $\delta_{2000}$ = $-$6:50:01.8,  
m$_{24}$ = 1.79 mag; and 2. $\alpha_{2000}$ =18:38:12.3, $\delta_{2000}$ = $-$6:52:01.5, m$_{24}$ = 4.44 mag) 
associated with the southern condensation are detected only in 24 $\mu$m image. 
It confirms that the stars are being formed in the southern condensation/IRDC. 
The star formation activity is seen in all the filaments and the clumps (cN1 and cS1) except clump cN2 (see Figures~\ref{fig4}a and~\ref{fig12}). 
The clump cN2 (peak N(H$_{2}$) $\sim$3.5 $\times$ 10$^{22}$ cm$^{-2}$) might be associated with the 
deeply embedded YSOs, which are not visible in the infrared regime.
\subsection{Distribution of H-band Polarization Vectors}
\label{subsec:pol}
Figure~\ref{fig13}a shows the distribution of H-band polarization vectors towards the molecular cloud traced 
by the integrated $^{13}$CO intensity map. 
The polarization vectors of 234 stars are shown in Figure~\ref{fig13}a (see Section~\ref{sec:obser} for more details). 
There are no H-band polarization detections towards the dense regions (e.g. clumps cN2, and cS1) in the complex. 
The H-band polarization detections are observed for a few sources located within previously known NIR cluster including the O5--O6 star in clump cN1. 
To explore the distribution of polarization data, we show mean polarization vectors in Figure~\ref{fig13}b. 
In order to obtain the mean polarization data, we divided the spatial area into 
10 $\times$ 10 equal divisions and computed a mean polarization value of H-band sources located inside each specific division.  
The length of a vector indicates the degree of polarization, while the inclination of a vector represents 
the polarization equatorial position angle. 
Following the standard grain alignment mechanisms, the polarization vectors of background stars reveal 
the sky-projected component of the magnetic field direction \citep{davis51}. 
Figure~\ref{fig14}a shows the GPS NIR color-color diagram of sources having H-band polarization detections. 
We find the reddened background stars and/or embedded stars with (J$-$H) $\geq$ 1.0, and the foreground sources with (J$-$H) $<$ 1.0.
Based on the above analysis, we find that the majority of stars appears behind W42 and traces the plane-of-the-sky projection of
the magnetic field in the W42 complex. 
The statistical distributions of the degree of polarization and the polarization equatorial position angles of 234 stars are shown in Figures~\ref{fig14}b and~\ref{fig14}c, respectively. 
We find that the degree of polarization value for a large population is $\sim$2--4\%. 
The starlight polarimetric data show an ordered plane-of-the-sky component of the magnetic field (position angle $\sim$15$\degr$) without additional magnetic field components. 
The overall polarization distribution is almost uniform towards the W42 molecular cloud. 
Note that the waist of the bipolar nebula has position angle of $\sim$15$\degr$ (see Figure~\ref{fig1}), which is very similar to the 
orientation of the magnetic field in the plane of the sky.  
\citet{jones04} carried out K-band polarimetric observations towards previously known NIR cluster including massive star, and 
suggested a uniform magnetic field geometry threading through the entire cluster. 
They found  a mean position angle of 18$\degr$ and a dispersion polarization angle of 12$\degr$ for the stars, which are 
located within 30$\arcsec$ of the O5--O6 star.  
The distribution of GPIPS H-band polarization vectors towards the NIR cluster is consistent with the results of \citet{jones04}. 
Note that the previous polarization study was restricted only towards the NIR cluster region, 
while the polarization study in this work is presented for a larger area towards the complex.

The exact position angle of the Galactic magnetic field is not known. 
However, \citet{heiles00} listed the optical polarimetry of stars in the surrounding sky of W42.
\citet{jones04} utilized the work of \citet{heiles00}, and suggested that 
the magnetic field direction in the diffuse ISM surrounding W42 lies in the Galactic plane 
at position angle of $\sim$28$\degr$. 

The plane-of-the-sky projection of the magnetic field (i.e. mean field) associated with W42 molecular 
cloud is close to the position angle of the Galactic magnetic field. 
It indicates that the mean field is not affected by self-gravity and/or turbulence 
during the cloud and core formation processes. 
Consequently, this result suggests that the Galactic magnetic field appears to be the prominent magnetic field 
even within the molecular cloud (including embedded H\,{\sc ii} region and filaments). 
Further discussion on the role of magnetic field is presented in Section~\ref{subsec:mag}.

On the eastern side of the bipolar lobe, we find a change in behavior of the polarimetric data (length and position angle) 
with respect to the polarization of stars located within previously known NIR cluster (see Figure~\ref{fig13}), 
where warm dust emission is dominated. 

In order to get an idea about magnetic field strength in W42, we estimated the plane-of-the-sky component, B$_{pos}$, following the equation given in 
\citet{chandrasekhar53}: 
\begin{equation}
B_{pos} = \Big{(} \frac{4}{3} \: \pi \: \rho \Big{)}^{0.5} \: \frac{\sigma_{v}}{\alpha}  \: \: \: \: \: [G],
\end{equation}
where $\rho$ is the volume mass density (in g cm$^{-3}$), $\sigma_{v}$ is the $^{13}$CO gas velocity dispersion (in cm s$^{-1}$), 
and $\alpha$ is the angular dispersion of the polarization vectors (in radians).
We fitted a single Gaussian to the integrated $^{13}$CO velocity profile and obtained $\sigma_{v}$ = 1.19 km s$^{-1}$. 
In this calculation, we use $\alpha$ = 12$\degr$ (mean value; as mentioned above) and $\rho$ $\sim$9.3 $\times$ 10$^{-21}$g cm$^{-3}$, 
which is computed using the H$_{2}$ number density ($\sim$2051 cm$^{-3}$; see Section~\ref{sec:dynage}) 
multiplied by molecular hydrogen's weight (2 $\times$ 1.00794 $\times$ 1.67 $\times$ 10$^{-24}$ gm) and a factor of 1.36 to account for helium 
and heavier elements. We find $B_{pos}$ $\sim$113.7 $\mu$G, which can be converted to magnetic 
pressure, $P_{mag}$ (= $B_{pos}^2(G)/8\pi$; dynes cm$^{-2}$), equal to $\sim$5.1 $\times$ 10$^{-10}$ dynes cm$^{-2}$. 
If we use a distance of 2.2 kpc to W42, then we obtain the H$_{2}$ number density $\sim$3541 cm$^{-3}$, $\rho$ $\sim$1.6 $\times$ 10$^{-20}$g cm$^{-3}$, 
$B_{pos}$ $\sim$148.5 $\mu$G, and $P_{mag}$ $\sim$8.8 $\times$ 10$^{-10}$ dynes cm$^{-2}$. 
These values are relatively higher than that derived at a distance of 3.8 kpc. 
The values of $\rho$, $P_{mag}$, and H$_{2}$ number density estimated at 3.8 kpc are found to be about 58\% of that derived at 2.2 kpc. 
However, the value of $B_{pos}$ calculated at 3.8 kpc is obtained to be about 76\% of value estimated at 2.2 kpc. 
Note that we do not have polarimetric data towards the densest regions (i.e. cN2 and cS1 clumps) in the complex, 
therefore this number can be taken as an indicative value for the complex. 
\section{Discussion}
\label{sec:data1}
\subsection{Feedback effect of the O-type star on the parental molecular cloud}
\label{subsec:feed}
\citet{deharveng10} used the multi-wavelength data (infrared, radio, and sub-mm) to 
explore the triggered star formation at the periphery of 102 MIR bubbles including the bipolar nebula, N39 (W42 region). 
They discussed the theoretical processes given in \citet{bodenheimer79} and \citet{fukuda00} and 
explained the presence of the bipolar morphology in N39 as due to the expansion of H\,{\sc ii} region in a filament.
They suggested that the ionized gas leaked out in two directions perpendicular to the dense regions.

The analysis of CO line kinematics provided the information of the distribution of gas and its morphological 
shape in W42 complex. 
The position-velocity analysis showed the signature of the expanding H\,{\sc ii} region (i.e. inverted C-like structures). 
The location of the massive star (O5--O6) appears to be associated with the warmest region ($\sim$30--36 K) compared to 
other surrounding regions ($\sim$20 and 25~K). The variation in the temperature inferred by the {\it Herschel} data is evident within the complex. 
A PDR region surrounding the H\,{\sc ii} region is traced by the H$_{2}$ (2.12 $\mu$m) and 8 $\mu$m emissions. 
Two cavities empty of molecular CO gas in the south-west and south-east directions indicated that the molecular gas is likely
to be eroded by the ionized gas. These cavities directly illustrate the interaction between the ionized gas of the H\,{\sc ii} region 
and the molecular environment in the complex. 
The W42 H\,{\sc ii} region seems to have eroded its parental molecular cloud in the southern side, thus giving rise to 
what are known as the northern and southern condensations. This argument is supported by the fact that the position angles of 
both the condensations being the same ($\sim$15$\degr$). The influence of UV photons on the southern 
condensation is also seen with the detection of the H$_{2}$ emission on its tip.  
A comparison of pressure contributions from different components associated with the massive star 
(i.e., P$_{HII}$, $P_{rad}$, and P$_{wind}$) suggests that the P$_{HII}$ dominates the other pressure components. 
The distribution of Class~II candidate YSOs \citep[average age $\sim$1--3 Myr;][]{evans09} is associated with the W42 molecular 
cloud (see Section~\ref{subsec:sfm}). 
The dynamical age for the H\,{\sc ii} region is estimated to be $\sim$0.32 Myr, which should be taken as an
indicative value. This value suggests that 
a single elongated structure was present prior to the formation of massive star. 
With the time, the impact of the H\,{\sc ii} region might have taken place within the cloud, leading to the subsequent 
formation of the two condensations.

\citet{dale13} have performed the smoothed particle hydrodynamics (SPH) numerical
simulations of the ionizing feedback effects from the O-type stars on the turbulent star-forming clouds.
They described the formation of the bipolar bubble-like structures in the parsec-scale, due to the ionizing feedback effects. 
Our findings also support the bipolar appearance of W42 complex as a result of the ionizing feedback from the  O5-O6 type star.

In the eastern side of the bipolar bubble, a change in the distribution of polarization data (values and angles) is found. 
A variation in the mean angular dispersion of the polarization vectors ($\alpha$) is also found towards the eastern lobe of the 
bipolar bubble ($\alpha$ $\sim$40$\degr$) compared to the main molecular cloud ($\alpha$ $\sim$12$\degr$). 
It is quite possible that the ionized gas front could sweep into the dust grains, causing a noticeable change in the polarization values and angles. 
It seems that the ionizing gas has expanded in the vicinity and is responsible for 
the observed structure of the gas distribution and of the polarization distribution. These observational features are consistent with 
the radiation-magnetohydrodynamic simulations of the expansion of H\,{\sc ii} region around an O star 
in a turbulent magnetized molecular cloud \citep{arthur11}. 
The authors pointed out that the presence of magnetic fields is vital for the morphology of small-scale features, such 
as globules and interstellar filaments. 
The results from the simulations also show that the expansion of H\,{\sc ii} region can influence the shape of the magnetic field lines.

All put together, our results show an imprint of the interaction between the H\,{\sc ii} region and its parental molecular cloud. 
\subsection{Role of magnetic field in W42 complex}
\label{subsec:mag}
The knowledge of the relative orientation of the molecular cloud and the mean field direction 
allows to infer the role of magnetic fields in the formation and evolution of the molecular cloud. 
In Section~\ref{subsec:pol}, we inferred that the position angles of molecular cloud (northern and southern condensations) 
and the starlight polarization angles are consistent with the Galactic magnetic field. Note that the observed magnetic 
field (plane-of-the-sky projection component) is uniform and is primarily parallel to the Galactic magnetic field. 
Therefore, it suggests the influence of the Galactic magnetic field in the evolution of the molecular cloud. 
Similar results were also obtained for a massive star-forming region G333.6$-$0.2 \citep{fujiyoshi01}. 
\subsection{Star formation in W42 complex}
\label{subsec:sfm}
The molecular material in the region (see Figure~\ref{fig1}) could be located at different distances. 
However, the $^{13}$CO line profile along the line of sight allowed us to trace the molecular cloud 
associated with W42 complex. In Section~\ref{subsec:surfden}, we studied YSO clusters associated with W42 
and also suggested the presence of a fraction of candidate YSOs located at larger distances.  
The distribution of clusters of YSOs (Class~I and Class~II) illustrates the star formation activity within W42 molecular cloud (including the filaments). 
In Section~\ref{sec:dynage}, we find that the parental gas has been affected by the ionized gas. 
The spatial locations of the YSO clusters in the molecular cloud indicate that the triggered star 
formation scenario could be applicable in W42 complex. 
One of the triggered star formation mechanisms suggests that the expanding H\,{\sc ii} 
region initiates the instability and helps in the collapse of a pre-existed dense clump in the molecular material \citep[e.g.][]{bertoldi89}. 
Gravitational instability within pre-existent condensations compressed by the H\,{\sc ii} region is unlikely prior to the dynamical age of the H\,{\sc ii} region.
The Class~I and Class~II YSOs have an average age of $\sim$0.44 Myr and $\sim$1--3 Myr \citep{evans09}, respectively.  
A relative comparison of these ages suggests that the age of Class~I YSOs is comparable to the dynamical age of the H\,{\sc ii} region ($\sim$0.32 Myr), 
while the age of Class~II YSOs is higher than the dynamical age of the H\,{\sc ii} region. Therefore, the evolved populations (i.e. Class~II YSOs) are 
unlikely to have been the product of triggered formation. 
A small fraction of Class~I candidate YSOs is identified in the northern and southern condensations, that might have been influenced by the H\,{\sc ii} region. 
As mentioned before, one should consider the dynamical timescale of the H\,{\sc ii} region with caution.

We notice that the different star formation processes have taken place in the northern and southern condensations. 
The southern condensation harbors the densest (peak N(H$_{2}$) $\sim$5 $\times$ 10$^{22}$ cm$^{-2}$), embedded (A$_{V}$ $\sim$53.5 mag), cold (20 K), 
and massive dust clump ($\sim$2564 $M_\odot$) in the W42 complex. The clump is associated with an embedded cluster of YSOs. 
Additionally, two embedded sources 
associated with the southern condensation are traced only in 24 $\mu$m image.
The NIR starlight polarimetric observations do not allow to obtain the magnetic field information within this particular dense clump.
Therefore, we can not discuss more about the role of magnetic field in the star formation process. 
However, a signature of gravitational instability is obtained using the virial mass ratio analysis (see Section~\ref{sec:dynage}). 
Consequently, it seems that the cluster of YSOs associated with the southern condensation is most-likely originated by gravitational instability. 

As mentioned earlier, in the northern condensation, at the heart of W42 complex, there appears a parsec-scale cavity-like structure, 
which encompasses different early evolutionary stages of massive star 
formation (i.e. B0V star, O5-O6 star, and W42-MME) (in two dimensional projection). 
The ionized cavity created by the UV photons of massive star(s) located in the NIR cluster, is not able to confine the ionized gas within itself, 
which is illustrated by the observed extended ionized emission seen in the 20 cm map. 
As mentioned earlier, the NIR cluster contains the O5-O6 star and 
two compact radio sources (i.e. G025.3824$-$00.1812 and G025.3809$-$00.1815). 
The line-of-sight velocity of the ionized gas in the W42 H\,{\sc ii} region \citep[$\sim$59.1 km\,s$^{-1}$;][]{lester85} is very similar to the 
velocity of the 6.7-GHz methanol maser \citep[$\sim$58.1~km\,s$^{-1}$;][]{szymczak12} and of 
the W42 molecular cloud \citep[58--69~km\,s$^{-1}$;][]{anderson09}. 
The velocity information confirms the physical association of molecular emission, ionized emission, 
and methanol maser emission. \citet{blum00} reported an extinction of A$_{V}$ $\sim$10 mag towards the 
NIR cluster using the average color of the brightest seven stars, including the spectroscopically 
identified O5-O6 star located within the cluster. The extinction of W42-MME was reported to 
be A$_{V}$ $>$ 45 mag \citep{jones04,dewangan15b}. 
Based on these different values of extinction in W42, \citet{jones04} suggested the presence of star 
formation activity behind the interface between the H\,{\sc ii} region and the molecular cloud.  
Based on our multi-wavelength data,  W42-MME appears to be a more deeply embedded source compared to 
the O and B type stars. Our analysis is also in agreement with the interpretation of \citet{jones04}.
However, all these sources seem to be located in the same complex. 

Additionally, a ``hub-filament" morphology or a filamentary system is seen in the northern condensation (see Section~\ref{subsec:temp}). 
We find that the cavity-like structure is located at the junction of the filaments (see filaments ``a, b, c, d, f, g, and h" in Figure~\ref{fig4}a). 
Note that the filaments have lower density compared to the hub/clump associated with the cavity-like structure (see Section~\ref{subsec:temp}). 
It suggests that the lower density filamentary structures interconnect at the high density region (see the zoomed-in view in Figure~\ref{fig5}). 
The spatial correlation of clusters of YSOs and the filaments is evident in Figure~\ref{fig12}. 
There exist in the literature similar observational evidences on other cloud complex available, such as Taurus, Ophiuchus, and Rosette  \citep[e.g.][]{myers09,schneider12}. 
Recently, \citet{schneider12} studied the Rosette Molecular Cloud using {\it Herschel} data and 
argued that the infrared clusters were preferentially found at the junction of filaments or filament mergers and their findings are 
consistent with the results obtained in the simulations of \citet{dale11}.  
The data presented in this work cannot throw light on the direct application of this scenario in the northern condensation. 
A detailed knowledge of the motion of the molecular material along the filaments is required to 
further investigate the role of filaments in the formation of the YSO clusters.
\section{Summary and Conclusions}
\label{sec:conc}
\begin{figure*}
\epsscale{1.0}
\plotone{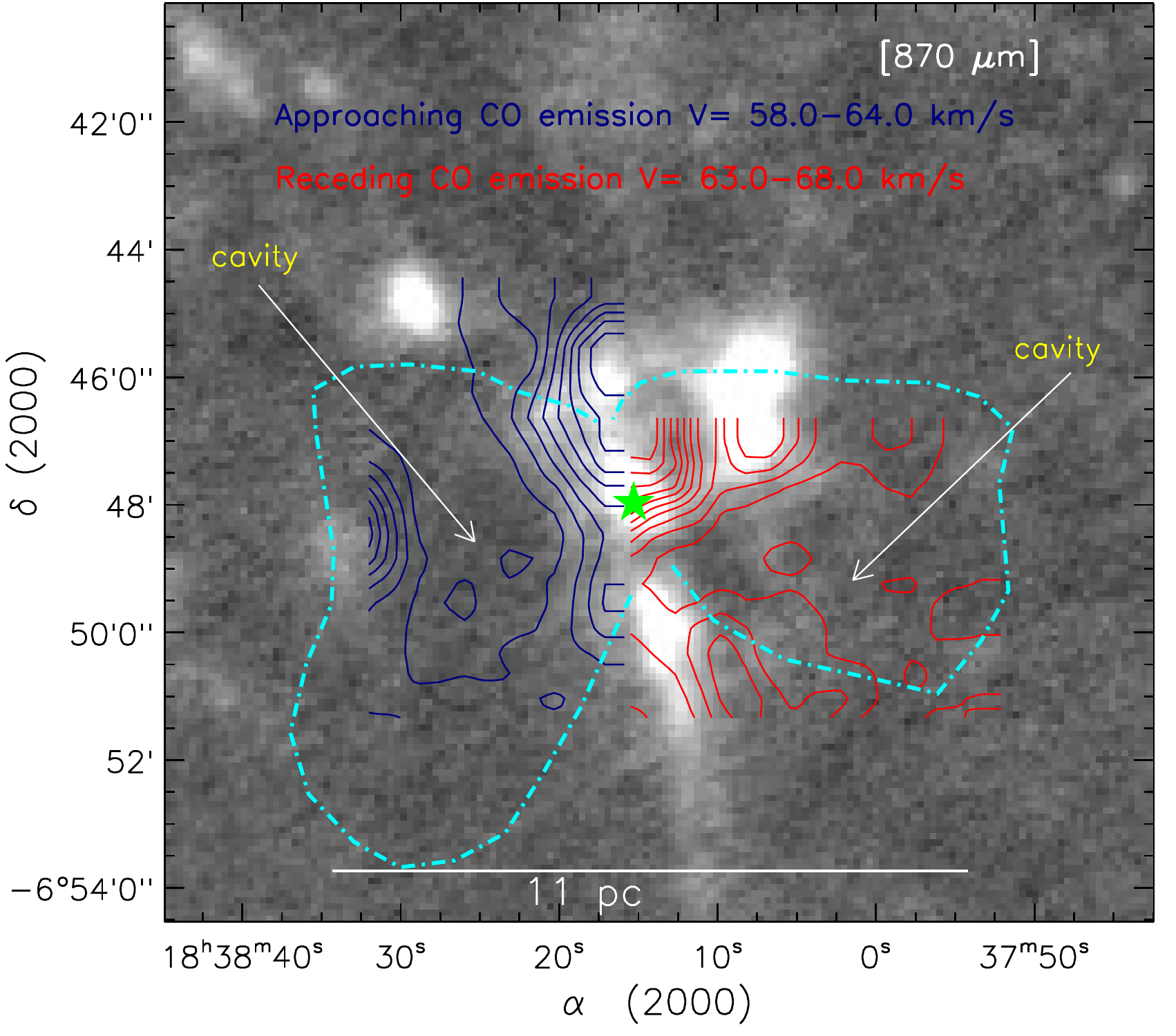}
\caption{\scriptsize Overlay of the integrated molecular emissions on the ATLASGAL 870 $\mu$m continuum image. 
A bipolar nebula is also highlighted similar to the one shown in Figure~\ref{fig1}. The star symbol indicates the location of an O5-O6 star. 
The empty molecular gas regions (i.e. cavities) are observed in the south-west and south-east 
directions with respect to the O5-O6 star (see figure~\ref{fig8a}). 
Cavities are traced at velocity ranges of 63--68~km\,s$^{-1}$ (red contours) and 58--64~km\,s$^{-1}$ (blue contours).
The CO red contours  are overlaid with levels of 
1.1, 3.2, 5.3, 7.5, 9.6, 11.7, 13.9, 16.0, 18.1, 20.2, 22.4, and 24.5 K km\,s$^{-1}$. 
The CO blue contours are drawn with levels of 1.1, 3.7, 6.2, 8.8, 11.5, 14.0, 16.6, 19.2, 21.8, and 24.4 K km\,s$^{-1}$. }
\label{fig9}
\end{figure*}

\begin{figure*}
\epsscale{1.0}
\plotone{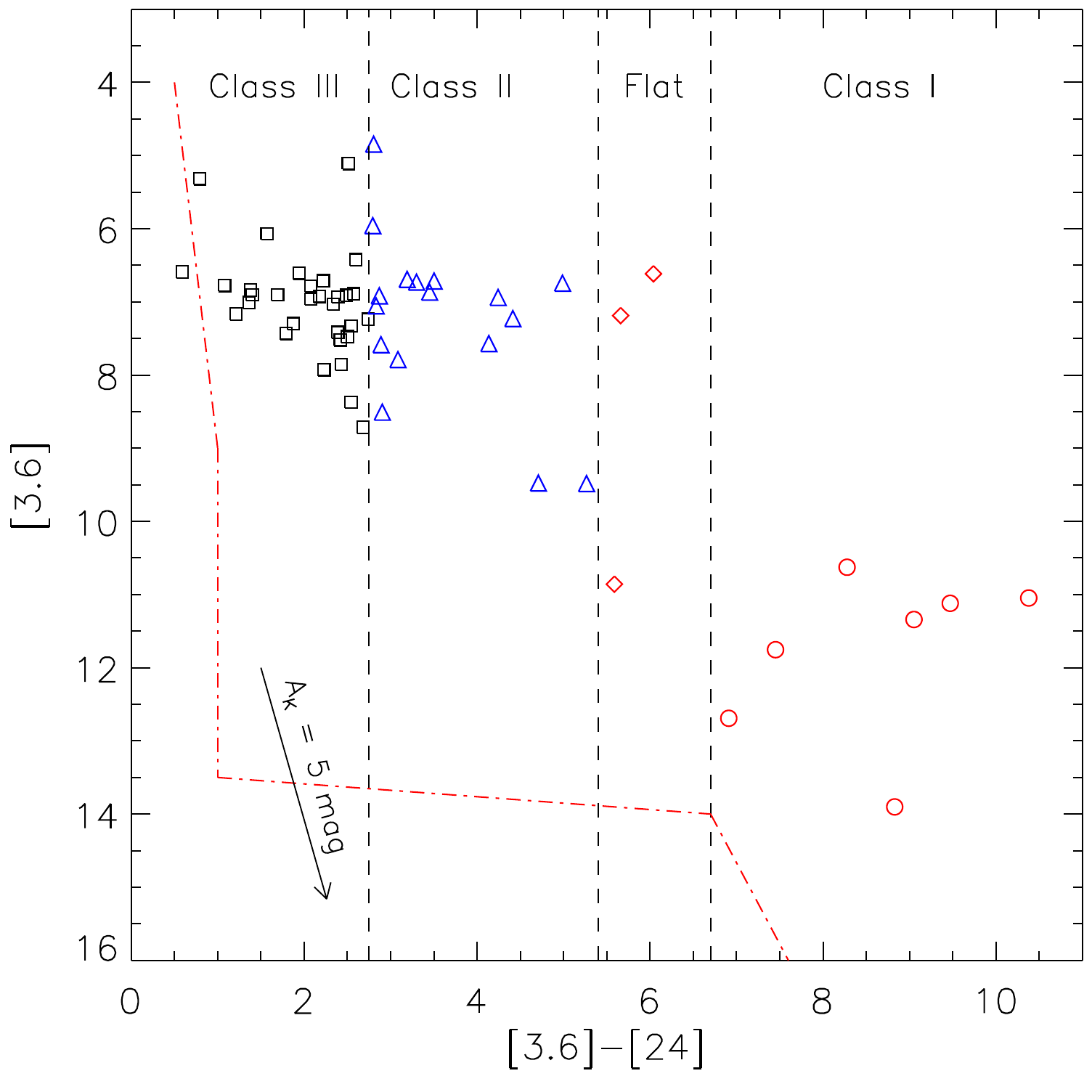}
\caption{\scriptsize Color-magnitude diagram ([3.6] $-$ [24] vs [3.6]) of sources detected in the IRAC and 
MIPSGAL bands (see the text for more details). Class~I, Class~II, Flat-spectrum, and Class~III 
sources are shown by open red circles, open blue triangles, open red diamonds, and open black squares, respectively. 
The dashed lines delineate the different spaces of these sources occupied in the color-magnitude diagram. 
The spaces of contaminated sources (galaxies and disk-less stars) and YSOs are separated by 
the dot-dashed lines \citep[see][for more details]{rebull11}. 
The arrow shows the extinction vector (A$_{K}$ = 5 mag) obtained using the average extinction law from \citet{flaherty07}.} 
\label{fig10a}
\end{figure*}

\begin{figure*}
\epsscale{1.0}
\plotone{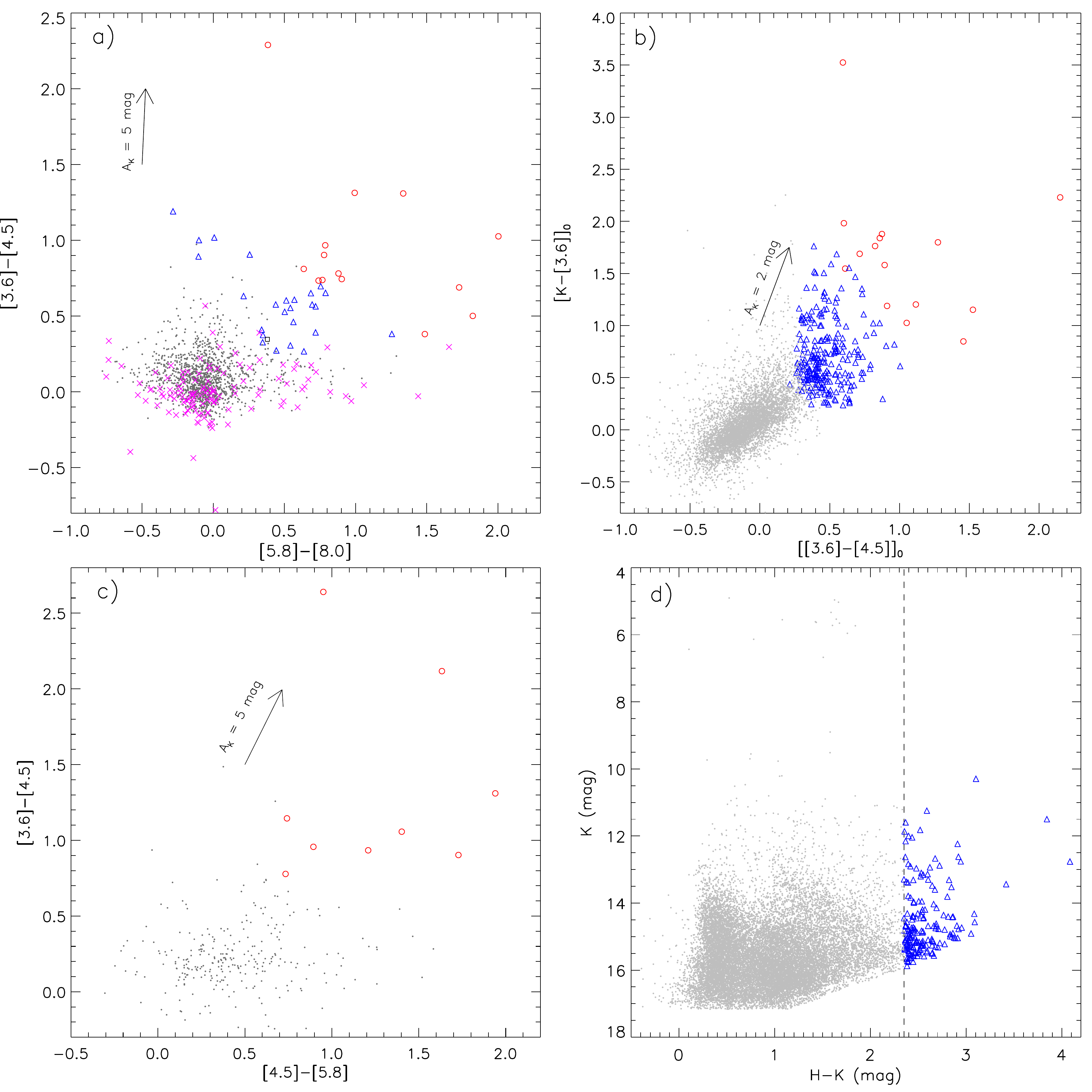}
\caption{\scriptsize a) Color-color diagram ([3.6]$-$[4.5] vs. [5.8]$-$[8.0]) using the {\it Spitzer}-IRAC four band detections. 
The ``$\times$'' symbols in magenta color show the identified PAH-emission-contaminated apertures in the region (see the text); 
b) The dereddened [K $-$ [3.6]]$_{0}$ $vs$ [[3.6]$-$[4.5]]$_{0}$ color-color diagram using WFCAM and IRAC data; 
c) Color-color diagram ([3.6]$-$[4.5] vs. [4.5]$-$[5.8]) of the sources detected in three IRAC bands, except 8.0 $\mu$m image; 
d) Color-magnitude diagram (H$-$K/K) of the sources detected in H and K bands. 
In all the panels, Class~I and Class~II candidate YSOs are shown by open red circles and open blue triangles, respectively. 
The dots in gray color represent the stars with only photospheric emissions (see the text for YSOs selection criteria). 
In the first three panels, the arrow shows the extinction vector obtained using the average extinction law from \citet{flaherty07}. } 
\label{fig10b}
\end{figure*}

\begin{figure*}
\epsscale{1.0}
\plotone{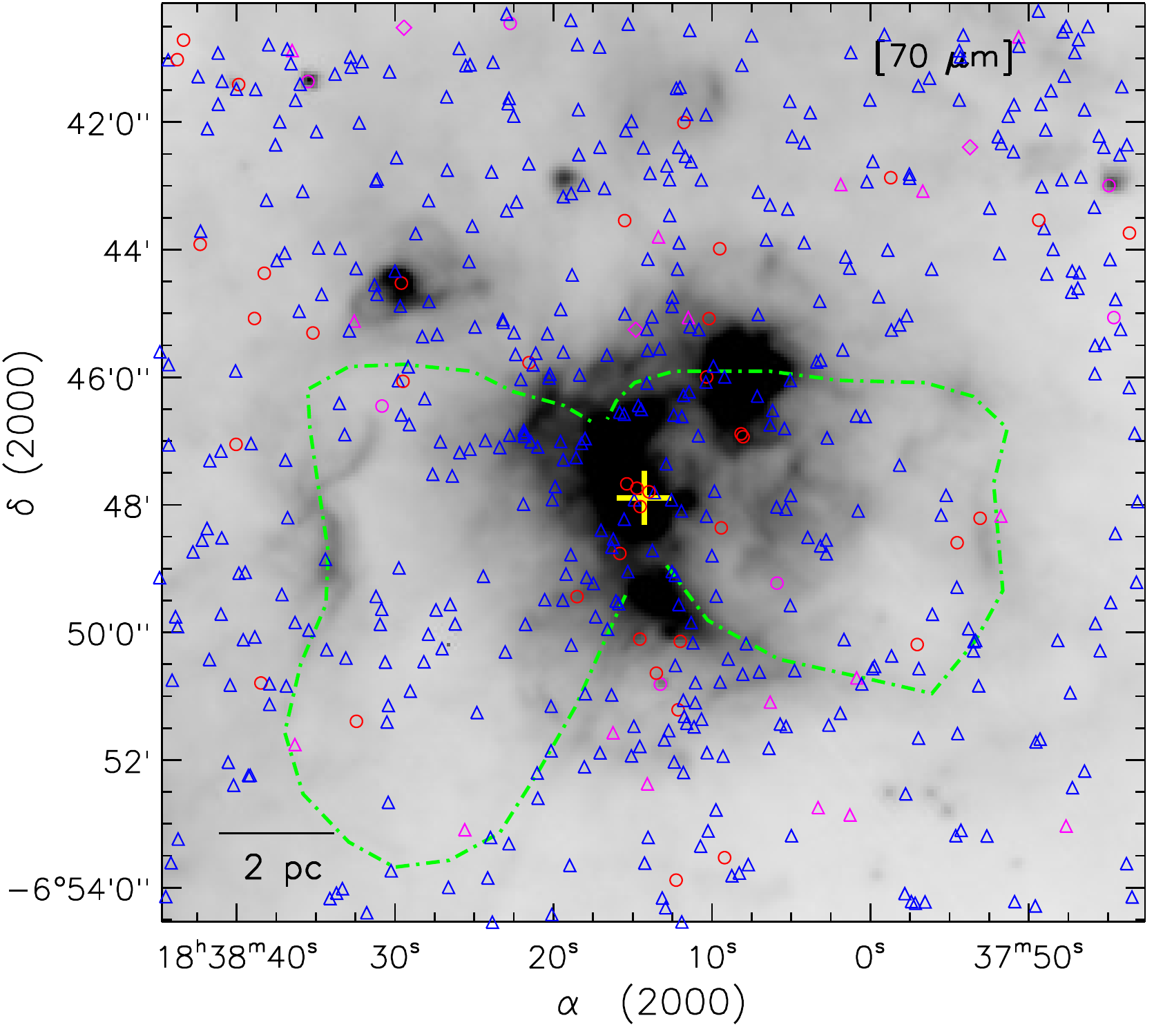}
\caption{\scriptsize The positions of Class~I (circles), Flat-spectrum (diamond), and Class~II (triangles) candidate YSOs identified within our selected region. 
The candidate YSOs identified using the IRAC-MIPSGAL scheme (see Figure~\ref{fig10a}) are shown by magenta color, 
while the candidate YSOs selected using the other schemes (see Figure~\ref{fig10b}) are highlighted in blue (Class~II) and red (Class~I) colors. 
The background map shows an inverted gray scale {\it Herschel} 70 $\mu$m continuum image. 
The position of IRAS 18355$-$0650 source is shown by a ``+" symbol. A bipolar nebula is also highlighted similar to the one shown in Figure~\ref{fig1}.} 
\label{fig11} 
\end{figure*}

\begin{figure*}
\epsscale{1.0}
\plotone{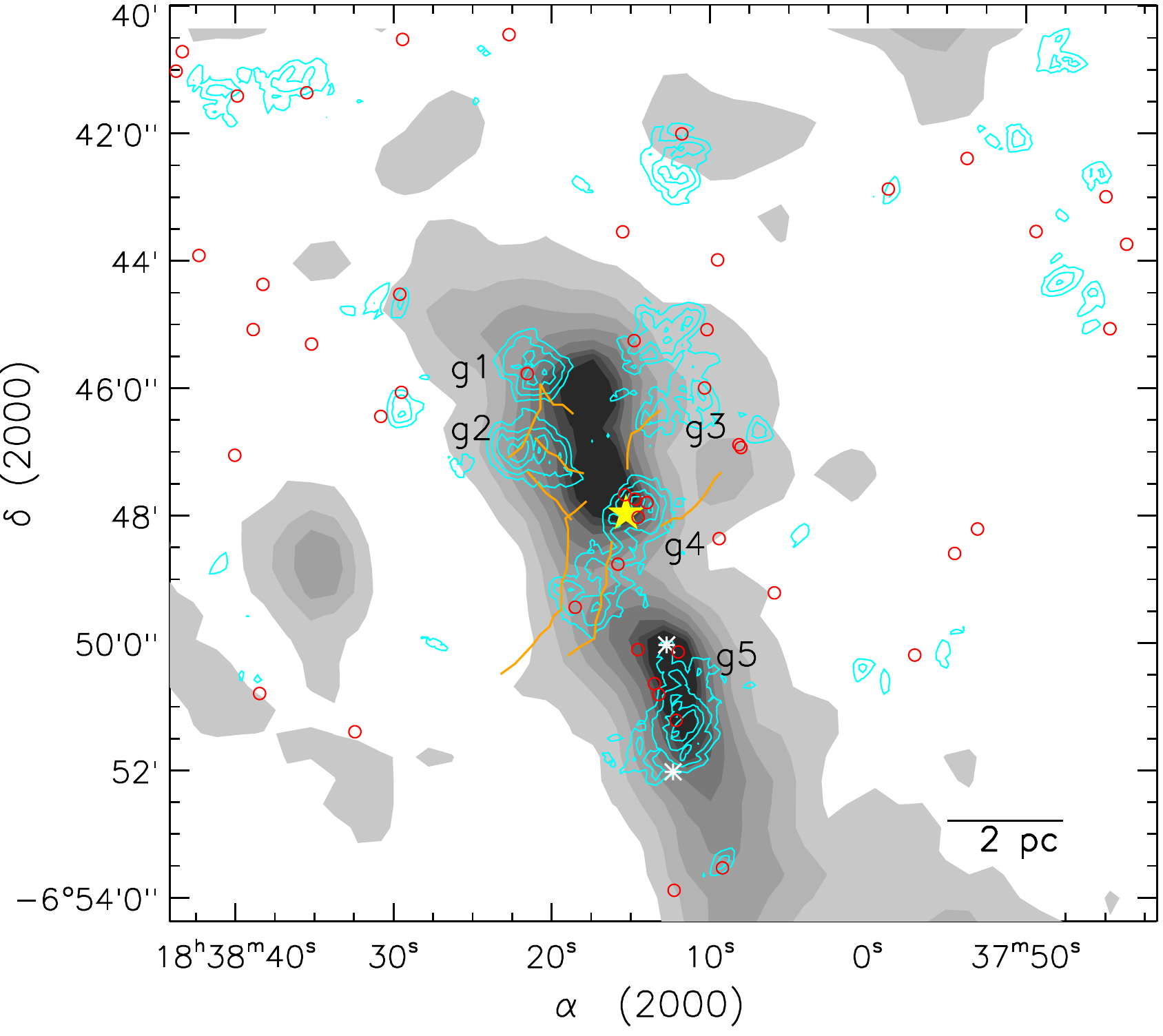}
\caption{\scriptsize Overlay of YSOs surface density contours on the integrated $^{13}$CO (J=1$-$0) emission 
contour map. The surface density contours are drawn at 3$\sigma$ (4 YSOs/pc$^{2}$, where 1$\sigma$=1.4 YSOs/pc$^{2}$), 
4$\sigma$ (6 YSOs/pc$^{2}$), 6$\sigma$ (8 YSOs/pc$^{2}$), and 9$\sigma$ (13 YSOs/pc$^{2}$), increasing from the outer to the inner regions.  
The background map is similar to the one shown in Figure~\ref{fig3}f. Class~I candidate YSOs are shown by open red circles. 
Five YSO clusters are labeled in the main molecular cloud. Two sources are marked by white asterisk symbols, which are detected only in 24 $\mu$m map. 
The {\it Herschel} filaments are overplotted by orange color curves (see Figure~\ref{fig4}a). 
The position of an O5-O6 star is marked with a filled star symbol.}
\label{fig12} 
\end{figure*}

\begin{figure*}
\epsscale{1.0}
\plotone{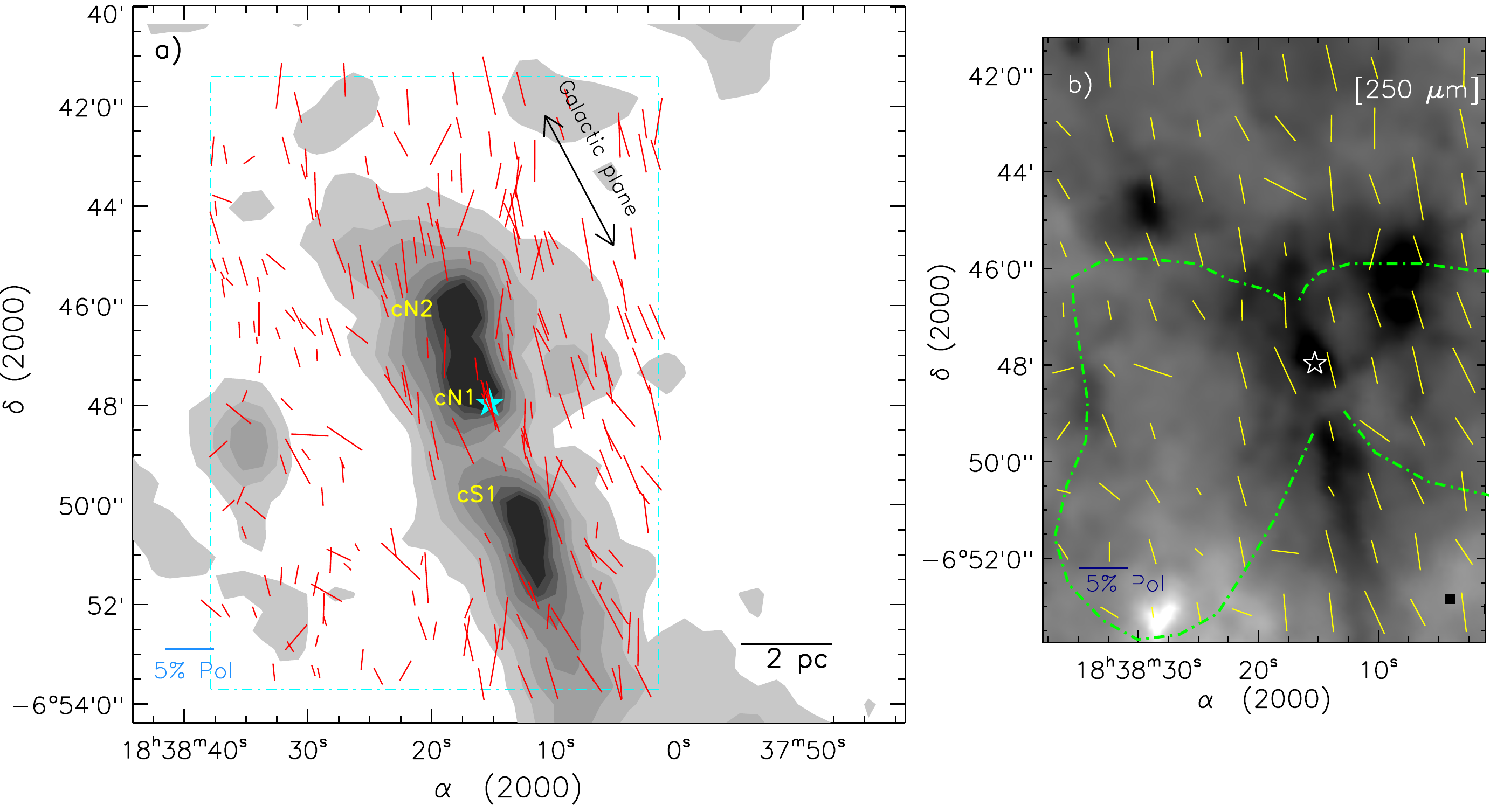}
\caption{\scriptsize a) Overlay of GPIPS H-band polarization vectors (in red color) of 234 stars with UF = 1 
and $P/\sigma_p \ge$ 2 on the integrated $^{13}$CO emission map. 
The background map is similar to the one shown in Figure~\ref{fig3}f. 
Vectors are drawn for the region marked by a dot-dashed box in figure, which are 
retrieved from the GPIPS fields GP0612 ($l$ = 25$\degr$.319, $b$ = $-$0$\degr$.240) and 
GP0626 ($l$ = 25$\degr$.447, $b$ = $-$0.$\degr$165). The length of each vector represents 
the degree of polarization (a reference vector of 5\% is shown in the lower left corner of the map). 
The orientations of the vectors represent the equatorial position angles of polarization, which show that 
the surrounding interstellar magnetic field lies in the Galactic plane at position angle of 28$\degr$. 
b) Overlay of mean polarization vectors on the inverted {\it Herschel} 250 gray scale map. 
The polarization spatial area is shown by a box in Figure~\ref{fig13}a. 
The mean polarization data are obtained by dividing the polarization spatial area into 10 $\times$ 10 equal 
divisions and, a mean polarization value of H-band sources is estimated inside each specific division. A bipolar nebula is 
also highlighted similar to the one shown in Figure~\ref{fig1}. 
In both the panels, the star symbol indicates the location of an O5-O6 star.}
\label{fig13}
\end{figure*}

\begin{figure*}
\epsscale{0.93}
\plotone{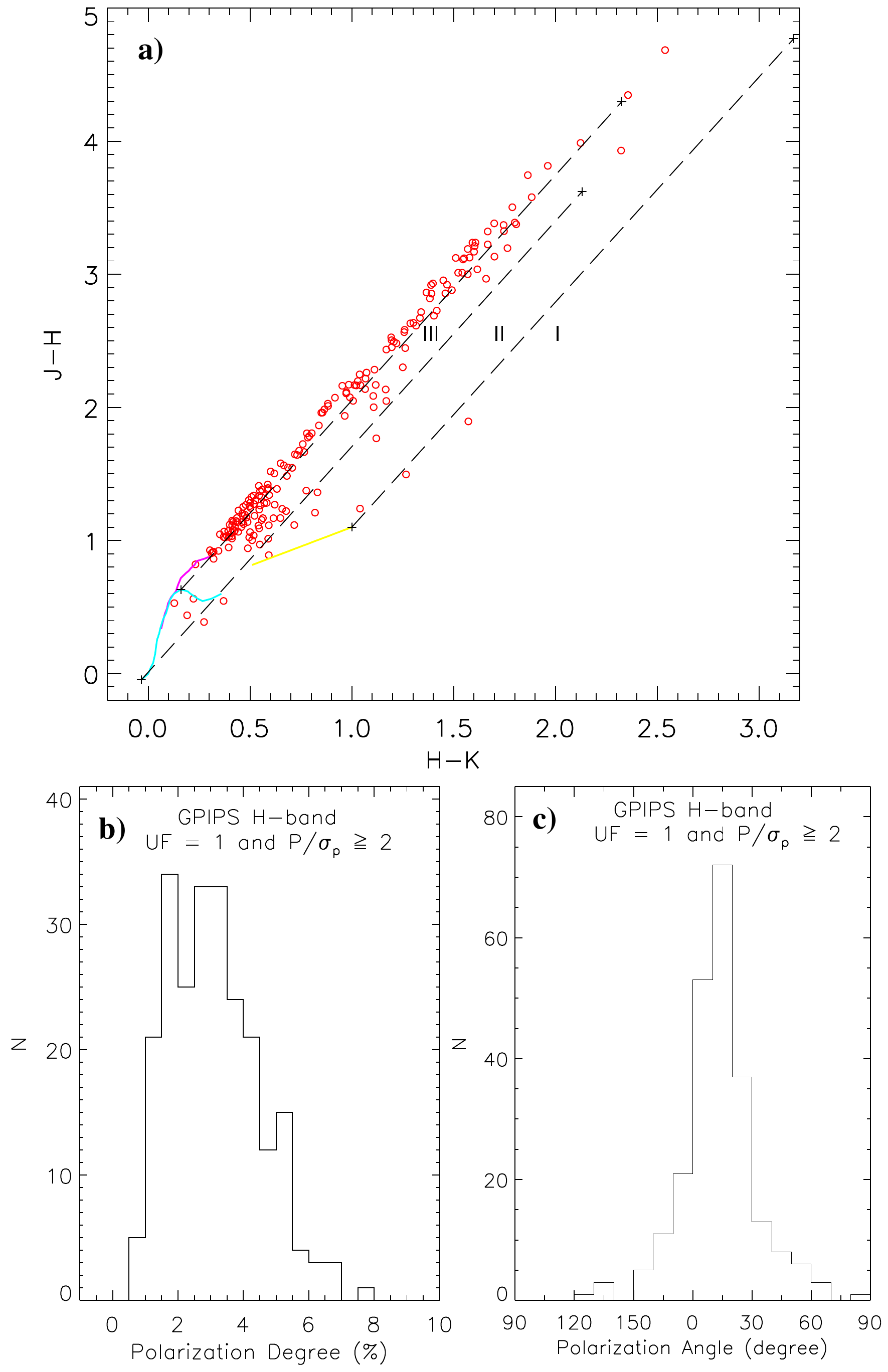}
\caption{\scriptsize a) GPS NIR color-color diagram (H$-$K vs J$-$H) of the sources having H-band polarization observations. The solid curves represent the unreddened locus of main sequence stars (cyan) and giants (magenta) \citep[from][]{bessell88}. 
The long-dashed straight lines show the extinction vectors with A$_{K}$ = 3 mag. 
The extinction vectors are derived from \citet{cohen81} extinction laws (A$_{J}$/A$_{K}$ = 2.94 and A$_{H}$/A$_{K}$ = 1.72 for California Institute of Technology (CIT) system). The color-color diagram is divided into three different subregions, namely ``I'', ``II'', and ``III''. 
The YSOs are identified in ``I'' and ``II'' subregions. Classical T Tauri (CTTS) locus (in CIT system) \citep{meyer97} is  
plotted as a solid yellow line. The loci of unreddened dwarf (Bessell \& Brett (BB) system), 
giant (BB-system), and colors are converted into CIT system using transformation 
equations provided by \citet{carpenter01}. 
b) The histograms of degree of polarization; and c) equatorial position angles of polarization obtained for 234 sources as mentioned in Figure~\ref{fig13}a. } 
\label{fig14} 
\end{figure*}

In this paper we have studied the physical environment, magnetic field, and stellar population in the W42 complex, 
using multi-wavelength data obtained from the publicly available surveys 
(i.e., MAGPIS, CORNISH, GRS, ATLASGAL, Hi-GAL, MIPSGAL, GLIMPSE, UWISH2, GPS, GPIPS, ESO-VLT, and 2MASS). 
We used high resolution 5 GHz radio continuum map and adaptive-optics NIR images to study the small scale environment of the most massive object. 
We utilized {\it Herschel} temperature and column density maps as well as $^{13}$CO (J=1$-$0) line kinematics to examine the 
physical conditions in the complex. We used different color-color and color-magnitude plots, as well as 
extinction map derived from the {\it Herschel} column density map, and the surface density analysis, to study the embedded YSOs in the complex. 
The main findings of our multi-wavelength analysis are the following:\\
$\bullet$ The largest structure in W42 complex is the bipolar nebula with an extension of $\sim$11 pc $\times$ 7 pc, as traced at 
wavelengths longer than 2 $\mu$m.\\ 
$\bullet$ {\it Herschel} dust emissions and $^{13}$CO gas show similar spatial morphology with 
two prominent condensations (i.e. northern and southern) along the waist axis of the bipolar nebula.\\ 
$\bullet$ The southern condensation is associated with the IRDC seen at 8.0 $\mu$m, which is seen as a prominent 
bright filament in emission at wavelengths longer than 70 $\mu$m. 
The kinematics of the CO gas towards the southern condensation suggest that the gas is 
moving away at a velocity of $\sim$1.5 km\,s$^{-1}$ with respect to the center of the bipolar nebula.\\ 
$\bullet$ The velocity of the ionized gas in the W42 H\,{\sc ii} region is very similar to the 
velocity of the 6.7-GHz methanol maser and of the W42 molecular cloud, confirming their physical association.\\
$\bullet$ The northern condensation at the heart of W42 complex contains a a parsec scale cavity-like structure which encompasses a B0V type object, a 
spectroscopically identified O5-O6 type object, and an infrared counterpart of the 6.7 GHz methanol 
maser (i.e. W42-MME) (in two dimensional projection), illustrating the presence of different evolutionary stages of massive star formation.\\
$\bullet$ The VLT/NACO adaptive-optics K and L$^{'}$ images resolved the O5-O6 type star into at least three 
point-like sources within a scale of 5000 AU.\\ 
$\bullet$ Two cavities of empty molecular gas (on scales of a few pc; see Figure~\ref{fig9}) are observed in the south-west 
and south-east directions with respect to the ionizing star, suggesting the ionized gas has probably escaped in these directions.\\
$\bullet$ The inverted C-like structures of molecular gas found in the position velocity maps suggest the signature of an expanding H\,{\sc ii} region.\\ 
$\bullet$ {\it Herschel} column density map traces two clumps in the northern condensation and one in the southern condensation.
The IRDC (southern condensation) appears to have a peak column density of $\sim$5 $\times$ 10$^{22}$ cm$^{-2}$, 
which corresponds to a visual extinction of A$_{V}$ $\sim$53.5 mag. 
The northern condensation shows the peak column densities of $\sim$3.6~$\times$~10$^{22}$ and $\sim$3.0~$\times$~10$^{22}$ cm$^{-2}$, 
which suggest visual extinctions of A$_{V}$ $\sim$38.5--32 mag.\\ 
$\bullet$ {\it Herschel} temperature map shows a variation in temperature within the complex.
The highest temperature ($\sim$36 K) is found towards the location of 5 GHz emissions. 
A PDR is traced with a temperature of $\sim$25 K. The southern condensation (i.e. IRDC) appears to have a temperature of 20~K.\\
$\bullet$ Parsec-scale filamentary structures are seen in the {\it Herschel} sub-mm continuum maps which appear to be radially 
pointed to the dense clump associated with massive stars, revealing a ``hub-filament" system. \\ 
$\bullet$ The distribution of H-band starlight polarization vectors shows a uniform magnetic field (plane-of-the-sky projection component) in the complex. 
The mean magnetic field associated with W42 complex is aligned along the Galactic magnetic field, suggesting the influence of the Galactic 
magnetic field lines on the evolution of the molecular cloud.\\
$\bullet$ 512 candidate YSOs are identified in the selected region, $\sim$40\% of which are present in clusters associated with the molecular cloud. \\
$\bullet$ The clusters of YSOs are distributed in the northern condensation including the {\it Herschel} filaments and the southern condensation. 
Additionally, the southern condensation harbors two embedded sources that are observed only in 24 $\mu$m image.\\ 
$\bullet$ In the northern condensation, the cluster of YSOs including massive star is located at the junction of the filaments.
In the southern condensation, a cluster of YSOs may have formed due to gravitational instability. 
High-resolution molecular line observations are necessary to further investigate the role of filaments in the formation of YSO clusters.
\acknowledgments
We thank the anonymous reviewers for a critical reading of the manuscript and several useful comments and 
suggestions, which greatly improved the scientific contents of the paper. 
This work is based on data obtained as part of the UKIRT Infrared Deep Sky Survey and UWISH2 survey. This publication 
made use of data products from the Two Micron All Sky Survey (a joint project of the University of Massachusetts and 
the Infrared Processing and Analysis Center / California Institute of Technology, funded by NASA and NSF), archival 
data obtained with the {\it Spitzer} Space Telescope (operated by the Jet Propulsion Laboratory, California Institute 
of Technology under a contract with NASA). 
This publication makes use of molecular line data from the Boston University-FCRAO Galactic
Ring Survey (GRS). The GRS is a joint project of Boston University and Five College Radio Astronomy Observatory, 
funded by the National Science Foundation (NSF) under grants AST-9800334, AST-0098562, and AST-0100793.  
This publication makes use of the Galactic Plane Infrared Polarization Survey (GPIPS). 
The GPIPS was conducted using the {\it Mimir} instrument, jointly developed at Boston University and Lowell Observatory
and supported by NASA, NSF, and the W.M. Keck Foundation. 
L.K.D. acknowledges the financial support provided by the CONACYT (Mexico) grant CB-2010-01-155142-G3 
for his postdoctoral fellowship.
The research is supported by CONACYT (M\'{e}xico) grants CB-2010-01-155142-G3 (PI. YDM) and CB-2012-01-1828-41 (PI. AL). 

\begin{deluxetable}{cccccccccccc}
%\tablewidth{0pc} 
\tablewidth{0pt} 
\tabletypesize{\scriptsize} 
%\setlength{\tabcolsep}{0.05in}
%\rotate
\tablecaption{Coordinates of some referred sources in W42, as highlighted in Figures~\ref{fig1}, \ref{fig2}, and \ref{fig3}. \label{tab1}}
\tablehead{ \colhead{ID} & \colhead{RA} & \colhead{Dec}\\
\colhead{} &  \colhead{[J2000]} & \colhead{[J2000]}}
\startdata
 O star 	                                             &  18:38:15.3	 &  -06:47:58.0\\ 								    
  G25.4NW region	                             &  18:38:08.3	 &  -06:45:48.7\\ 
 6.7-GHz methanol maser	                         &  18:38:14.5      &  -06:48:02.0\\ 
 G025.3809-00.1815   	                            &  18:38:15.0	 &  -06:48:01.2\\ 
 G025.3824-00.1812  	                            &  18:38:15.3	 &  -06:47:52.6\\ 
 cN1 condensation	                             &  18:38:17.3	 &  -06:47:24.3\\ 								    
 cN2 condensation 	                             &  18:38:17.3	 &  -06:47:09.8\\ 								    
 cS1 condensation 	                             &  18:38:11.6	 &  -06:50:40.4\\ 								    
\enddata 
\end{deluxetable}

\begin{appendix}
\section{SQL conditions to select reliable point sources from the GPS catalog in W42}\label{app:cond} 
\noindent{\textbf{For all {\it JHK}:}} {\sc select} ra, dec,
jAperMag3, jAperMag3Err, hAperMag3, hAperMag3Err, k\_1AperMag3,
k\_1AperMag3Err, pStar {\sc from} gpsSource {\sc where} $ra$ {\sc
  between} 279.42859 {\sc and} 279.68706 {\sc and} $dec$ {\sc between}
-6.9096591 {\sc and} -6.6704976 {\sc and} mergedClass != 0 {\sc and}
(PriOrSec=0 {\sc or} PriOrSec=framesetID) {\sc and} pstar $>$ 0.90
{\sc and} jppErrbits $<$ 256 {\sc and} hppErrbits $<$ 256 {\sc and}
k\_1ppErrbits $<$ 256 {\sc and} abs(jAperMag3Err) $<$ 0.1 {\sc and}
abs(hAperMag3Err) $<$ 0.1 {\sc and} abs(k\_1AperMag3Err) $<$ 0.1\\
\noindent{\textbf{For only {\it HK}:}} {\sc select} ra, dec,
jAperMag3, jAperMag3Err, hAperMag3, hAperMag3Err, k\_1AperMag3,
k\_1AperMag3Err, pStar {\sc from} gpsSource {\sc where} $ra$ {\sc
  between} 279.42859 {\sc and} 279.68706 {\sc and} $dec$ {\sc between}
-6.9096591 {\sc and} -6.6704976 {\sc and} mergedClass != 0 {\sc and}
(PriOrSec=0 {\sc or} PriOrSec=framesetID) {\sc and} pstar $>$ 0.90
{\sc and} hppErrbits $<$ 256 {\sc and} k\_1ppErrbits $<$ 256 {\sc and}
abs(hAperMag3Err) $<$ 0.1 {\sc and} abs(k\_1AperMag3Err) $<$ 0.1

\end{appendix}


\begin{thebibliography}{}

\bibitem[Ai et al.(2013)]{ai13}	
Ai, M., Zhu, M., Xiao, Li, Su, Hong-Quan 2013, RAA, 13, 935

\bibitem[Anderson et al.(2009)]{anderson09}
Anderson, L.~D., Bania, T.~M., Jackson, J.~M., et al. 2009, ApJS, 181, 255

\bibitem[Arce et al.(2011)]{arce11}
Arce, H.~G., Borkin, M.~A., Goodman, A.~A., Pineda, J.~E.,\& Beaumont, C.~N. 2011, ApJ, 742, 105

\bibitem[Arthur et al.(2011)]{arthur11}
Arthur, S.~J., Henney, W.~J., Mellema, G., \& Colle, F.~De 2011, MNRAS, 414, 1747

\bibitem[Beaumont \& Williams(2010)]{beaumont10}
Beaumont, C.~N., \& Williams, J.~P. 2010, ApJ, 709, 791

\bibitem[Benjamin et al.(2003)]{benjamin03}
Benjamin, R.~A.,Churchwell, E., Babler, B.~L., et al. 2003, PASP, 115, 953

\bibitem[Bertoldi(1989)]{bertoldi89}
Bertoldi, F. 1989, ApJ, 346, 735

\bibitem[Bessell \& Brett(1988)]{bessell88}
Bessell, M.~S., \& Brett J.~M. 1988, PASP, 100, 1134

\bibitem[Blum et al.(2000)]{blum00}
Blum, R.~D., Conti, P.~S., \& Damineli, A. 2000, AJ, 119 1860

\bibitem[Bodenheimer et al.(1979)]{bodenheimer79}
Bodenheimer, P., Tenorio-Tagle, G., \& Yorke, H.~W. 1979, ApJ, 233, 85

\bibitem[Bohlin et al.(2000)]{bohlin78}
Bohlin, R.~C., Savage, B.~D., \& Drake, J.~F. 1978, ApJ, 224, 13233

\bibitem[Bressert et al.(2012)]{bressert12}
Bressert, E., Ginsburg, A., Bally, J., Battersby, C., Longmore, S., \& Testi, L. 2012, ApJ, 758, 28

\bibitem[Carpenter(2001)]{carpenter01}
Carpenter, J.~M 2001, AJ, 121, 2851

\bibitem[Carey et al.(2005)]{carey05}
Carey, S. J., Noriega-Crespo, A., Price, S.~D., et al. 2005, BAAS, 37, 1252

\bibitem[Casali et al.(2007)]{casali07}
Casali, M., Adamson, A., Alves de Oliveira, C., et al. 2007, A\&A, 467, 777

\bibitem[Casertano \& Hut(1985)]{casertano85}
Casertano, S., \& Hut P. 1985, ApJ, 298, 80

\bibitem[Chandrasekhar \& Fermi(1953)]{chandrasekhar53}
Chandrasekhar, S., \& Fermi, E. 1953, ApJ, 118, 113

\bibitem[Chavarr\'ia et al.(2008)]{chavarria08}
Chavarr\'ia, L.~A., Allen, L.~E., Hora, J.~L., et al. 2008, ApJ, 682, 445

\bibitem[Churchwell et al.(2006)]{churchwell06}
Churchwell, E., Povich, M.~S., Allen, D., et al. 2006, ApJ, 649, 759

\bibitem[Clemens et al.(2012)]{clemens12}
Clemens, D.~P., Pavel, M.~D., \& Cashman, L.~R. 2012, ApJS, 200, 21

\bibitem[Cohen et al.(1981)]{cohen81}
Cohen, J.~G., Persson, S.~E., Elias, J.~H., \& Frogel, J.~A. 1981, ApJ, 249, 481

\bibitem[Dale \& Bonnell(2011)]{dale11}
Dale, J.~E.,  \& Bonnell, I.~A. 2011, MNRAS, 414, 321

\bibitem[Dale et al.(2013)]{dale13}
Dale, J.~E., Ercolano, B., \& Bonnell, I.~A. 2013, MNRAS, 431, 1062

\bibitem[Davis \& Greenstein(1951)]{davis51}
Davis, L., Jr., \& Greenstein, J. L. 1951, ApJ, 114, 206

\bibitem[Deharveng et al.(2010)]{deharveng10}
Deharveng, L., Schuller, F., Anderson, L.~D., et al. 2010, A\&A, 523, 6

\bibitem[de Jager et al.(1988)]{dejager88}	
de Jager, C., Nieuwenhuijzen, H.,\&  van der Hucht, K. A. 1988, A\&AS, 72,259

\bibitem[Dewangan \& Anandarao(2011)]{dewangan11}
Dewangan, L.~K., \& Anandarao, B.~G 2011, MNRAS, 414, 1526

\bibitem[Dewangan et al.(2012)]{dewangan12}
Dewangan, L.~K., Ojha, D.~K., Anandarao, B.~G., Ghosh, S.~K., \& Chakraborti, S. 2012, ApJ, 756, 151

\bibitem[Dewangan et al.(2015a)]{dewangan15}
Dewangan, L.~K., Ojha, D.~K., Grave, J.~M.~C., \& Mallick, K.~K. 2015a, MNRAS, 446, 2640

\bibitem[Dewangan et al.(2015b)]{dewangan15b}
Dewangan, L. K., Mayya, Y.~D., Luna, A., \& Ojha, D. K. 2015b, ApJ, 803, 100

\bibitem[Duch\^{e}ne \& Kraus(2013)]{duchene13}
Duch\^{e}ne, G., \& Kraus, A., ARA\&A 2013, 51, 269

\bibitem[Dye et al.(2006)]{dye06}
Dye, S., Warren, S.~J., Hambly, N.~C., et al. 2006, MNRAS, 372, 1227

\bibitem[Dyson \& Williams(1980)]{dyson80}
Dyson, J.~E., \& Williams, D.~A. 1980, Physics of the interstellar medium, New York, Halsted Press, 204 p

\bibitem[Evans et al.(2009)]{evans09}
Evans, N.~J., II, Dunham, M.~M., J\o{}rgensen, J.~K., et al. 2009, ApJS, 181, 321

\bibitem[Fazio et al.(2004)]{fazio04}
Fazio, G.~G., Hora, J.~L., Allen, L.~E., et al. 2004, ApJS, 154, 10

\bibitem[Flaherty et al.(2007)]{flaherty07}
Flaherty, K.~M., Pipher, J.~L., Megeath, S.~T., et al. 2007, ApJ, 663, 1069

\bibitem[Froebrich et al.(2011)]{froebrich11}
Froebrich, D., Davis, C.~J., Ioannidis, G., et al. 2011, MNRAS, 413, 480

\bibitem[Fujiyoshi et al.(2001)]{fujiyoshi01}
Fujiyoshi, T., Smith, C.~H., Wright, C.~M., et al. 2001, MNRAS, 327, 233

\bibitem[Fukuda et al.(2000)]{fukuda00}
Fukuda, N., \& Hanawa, T. 2000, ApJ, 533, 911

\bibitem[Garay et al.(1993)]{garay93} 
Garay, G., Rodriguez, L.~F., Moran, J.~M., \& Churchwell, E. 1993, ApJ, 418, 368

\bibitem[Getman et al.(2007)]{getman07} 
Getman, K.~V., Feigelson, E.~D., Garmire,G., Broos, P., \& Wang, J. 2007, ApJ, 654, 316 

\bibitem[Goto et al.(2006)]{goto06} 
Goto, M., Stecklum, B., Linz, H., Feldt, M., Henning, Th., Pascucci, I., \& Usuda, T. 2006, ApJ, 649, 299

\bibitem[Griffin et al.(2010)]{griffin10} 
Griffin, M.~J., Abergel, A., Abreu, A, et al. 2010, A\&A, 518L, 3

\bibitem[Guieu et al.(2010)]{guieu10}
Guieu, S., Rebull, L.~M., Stauffer, J.~R., et al. 2010, ApJ, 720, 46

\bibitem[Gutermuth et al.(2009)]{gutermuth09}
Gutermuth, R.~A., Megeath, S.~T., Myers, P.~C., et al. 2009, ApJS, 184, 18

\bibitem[Harper-Clark \& Murray(2009)]{harper09} 
Harper-Clark, E., \& Murray, N. 2009, ApJ, 693, 1696

\bibitem[Hartmann et al.(2005)]{hartmann05} 
Hartmann, L., Megeath, S.~T., Allen, L., et al. 2005, ApJ, 629, 881

\bibitem[Heiles(2000)]{heiles00}
Heiles, C. 2000, AJ, 119, 923

\bibitem[Helfand et al.(2006)]{helfand06}
Helfand, D.~J., Becker, R.~H., White, R.~L., Fallon, A., \& Tuttle, S. 2006, AJ, 131, 2525 

\bibitem[Hildebrand(1983)]{hildebrand83} 
Hildebrand, R.~H. 1983, Quarterly Journal of the RAS, 24, 267

\bibitem[Hoare et al.(2012)]{hoare12}
Hoare, M.~G., Purcell, C.~R., Churchwell, E.~B., et al. 2012, PASP, 124, 939 

\bibitem[Hodgkin et al.(2009)]{hodgkin09}
Hodgkin, S.~T., Irwin, M.~J., Hewett, P.~C., \& Warren, S.~J. 2009, MNRAS, 394, 675

\bibitem[Jackson et al.(2006)]{jackson06} 
Jackson, J.~M., Rathborne, J.~M., Shah, R.~Y., et al. 2006, ApJS, 163, 145

\bibitem[Jones et al.(2004)]{jones04} 
Jones, T.~J., Woodward, C.~E., \& Kelley, M.~S. 2004, ApJ, 128, 2448

\bibitem[Kauffmann et al.(2008)]{kauffmann08}
Kauffmann, J., Bertoldi, F., Bourke, T.~L., Evans, II, N.~J.,\&  Lee, C.~W. 2008, ApJ, 487, 993

\bibitem[Kwan(1997)]{kwan97} 
Kwan, J. 1997, ApJ, 489, 284

\bibitem[Lada et al.(2006)]{lada06}
Lada, C.~J., Muench, A.~A., Luhman, K.~L., et al. 2006, AJ, 131, 1574

\bibitem[Lawrence et al.(2007)]{lawrence07}
Lawrence, A., Warren, S.~J., Almaini, O., et al. 2007, MNRAS, 379, 1599

\bibitem[Lee et al.(2014)]{lee14}
Lee, J.~J., Koo,  B.~C., Lee, Y.~H., et al. 2014, MNRAS, 443, 2650

\bibitem[Lenzen et al.(2003)]{lenzen03}
Lenzen, R., Hartung, M., Brandner, W., et al. 2003, Proc. SPIE, 4841, 944

\bibitem[Lester et al.(1985)]{lester85}
Lester, D.~F., Dinerstein, H.~L., Werner, M.~W., Harvey, P.~M., Evans, N.~J., \& Brown, R.~L. 1985, AJ, 296, 565

\bibitem[Lucas et al.(2008)]{lucas08}
Lucas, P.~W., Hoare, M.~G., Longmore, A., et al. 2008, MNRAS, 391, 1281

\bibitem[MacLaren et al.(1988)]{maclaren88}
MacLaren, I., Richardson, K.~M., \& Wolfendale, A.~W. 1988, ApJ, 333, 821

\bibitem[Mallick et al.(2015)]{mallick15}
Mallick, K.~K., Ojha, D.~K., Tamura, M., Linz, H., Samal, M.~R., \& Ghosh, S.~K. 2015, MNRAS, 447, 2307

\bibitem[Martins et al.(2005)]{martins05}
Martins, F., Schaerer, D., \& Hillier, D.~J. 2005, A\&A, 436, 1049

\bibitem[Matsakis et al.(1976)]{matsakis76}
Matsakis, D.~N., Evans, N.~J., II, Sato, T., \& Zuckerman, B. 1976, AJ, 81, 172

\bibitem[Meyer et al.(1997)]{meyer97} 
Meyer, M.~R., Calvet, N., \& Hillenbrand, L.~A. 1997, AJ, 114, 288

\bibitem[Molinari et al.(2010)]{molinari10}
Molinari, S., Swinyard, B., Bally, J., et al. 2010, A\&A, 518, L100

\bibitem[Myers (2009)]{myers09} 
Myers, P.~C. 2009, ApJ, 700, 1609

\bibitem[Ott(2010)]{ott10}
Ott, S. 2010, in Astronomical Society of the Pacic Conference
Series, Vol. 434, Astronomical Data Analysis Software and
Systems XIX, ed. Y. Mizumoto, K.-I. Morita, \& M. Ohishi, 139

\bibitem[Poglitsch et al.(2010)]{poglitsch10}	
Poglitsch, A., Waelkens, C., Geis, N., et al. 2010, A\&A, 518L, 2

\bibitem[Prinja et al.(1990)]{prinja90}	
Prinja, R.~K., Barlow, M.~J., \& Howarth, I.~D. 1990, ApJ, 361, 607

\bibitem[Purcell et al.(2013)]{purcell13}	
Purcell, C.~R., Hoare, M.~G., Cotton, W.~D., et al. 2013, ApJS, 205, 1

\bibitem[Quireza et al.(2006)]{quireza06} 
Quireza, C, Rood, R.~T., Balser, D.~S., \& Bania, T.~M. 2006, ApJS, 165, 338

\bibitem[Rebull et al.(2011)]{rebull11}
Rebull, L.~M., Guieu, S., Stauffer, J.~R., et al. 2011, ApJS, 193, 25

\bibitem[Rousset et al.(2003)]{rousset03}
Rousset, G., Lacombe, F., Puget, P., et al. 2003, Proc. SPIE, 4839, 140

\bibitem[Samal et al.(2015)]{samal15}
Samal, M.~R., Ojha, D.~K., Jose, J., et al. 2015,  in A\&A press, arXiv150309037

\bibitem[Schneider et al.(2012)]{schneider12}
Schneider, N., Csengeri, T., Hennemann, M., et al. 2012, A\&A, 540, L11

\bibitem[Schuller et al.(2009)]{schuller09}
Schuller, F., Menten, K.~M., Contreras, Y., et al. 2009, A\&A, 504, 415

\bibitem[Shinn et al.(2014)]{shinn14}
Shinn, J.~H., Kim, K.~T., Lee, J.~J., et al. 2014, ApJS, 214, 11

\bibitem[Skrutskie et al.(2006)]{skrutskie06}
Skrutskie, M.~F., Cutri, R.~M., Stiening, R., et al. 2006, AJ, 131, 1163

\bibitem[Smith et al.(2002)]{smith02}
Smith, L.~J., Norris, R.~P.~F., \& Crowther, P.~A. 2002, MNRAS, 337, 1309

\bibitem[Szymczak et al.(2012)]{szymczak12} 	
Szymczak, M., Wolak, P., Bartkiewicz, A., \& Borkowski, K.~M. 2012, AN, 333, 634

\bibitem[Wienen et al.(2012)]{wienen12} 
Wienen, M., Wyrowski, F., Schuller, F., et al. 2012, A\&A, 544, 146

\bibitem[Williams et al.(1994)]{williams94} 
Williams, J. P., de Geus, E. J., \& Blitz, L. 1994, ApJ, 428, 693

\bibitem[Woodward et al.(1985)]{woodward85}
Woodward, C.~E., Helfer, H.~L., \& Pipher, J.~L. 1985, A\&A, 147, 84

\end{thebibliography}
\end{document}